\documentclass[conf]{new-aiaa}
\usepackage[utf8]{inputenc}
\usepackage{graphicx}
\usepackage{amsmath}
\usepackage[version=4]{mhchem}
\usepackage{siunitx}
\usepackage{bm}
\usepackage{longtable,tabularx}
\usepackage{adjustbox,lipsum}
\usepackage{subcaption}
\usepackage{xcolor}
\title{How machine learning can help the design and analysis of composite materials and structures?}

\author{Xin Liu\footnote{Assistant Professor, Department of Industrial, Manufacturing, and Systems Engineering.}} \affil{The University of Texas at Arlington, Arlington, TX 76019}
\author{Su Tian\footnote{Graduate Assistant, School of Aeronautics and Astronautics.} and Fei Tao\footnote{Graduate Assistant, School of Aeronautics and Astronautics.} and Haodong Du\footnote{Graduate Assistant, School of Aeronautics and Astronautics.} and Wenbin Yu\footnote{Professor, School of Aeronautics and Astronautics, AIAA Associate Fellow.}}
\affil{Purdue University, West Lafayette, IN 47906}

\begin{document}

\maketitle

\begin{abstract}
Machine learning models are increasingly used in many engineering fields thanks to the widespread digital data, growing computing power, and advanced algorithms. Artificial neural networks (ANN) is the most popular machine learning model in recent years. Although many ANN models have been used in the design and analysis of composite materials and structures, there are still some unsolved issues that hinder the acceptance of ANN models in the practical design and analysis of composite materials and structures. Moreover, the emerging machine learning techniques are posting new opportunities and challenges in the data-based design paradigm. This paper aims to give a state-of-the-art literature review of ANN models in the nonlinear constitutive modeling, multiscale surrogate modeling, and design optimization of composite materials and structures. This review has been designed to focus on the discussion of the general frameworks and benefits of ANN models to the above problems. Moreover, challenges and opportunities in each key problem are identified and discussed. This paper is expected to open the discussion of future research scope and new directions to enable efficient, robust, and accurate data-driven design and analysis of composite materials and structures.
\end{abstract}



\section{Motivation}
\lettrine{M}{achine} learning models are rapidly infiltrating many engineering fields due to its success in speech recognition, image recognition, language processing, etc. With the unprecedented growing data from experiments and computer simulations, rapid increasing computing power, and emerging advanced algorithms, one should expect unforeseeable and revolutionary impacts across nearly the entire domain of design and analysis of composite materials and structures over the next two decades \cite{dimiduk2018perspectives}. Recently, Peng at al. \cite{peng2020multiscale} reviewed the state-of-the-art work of machine learning in multiscale modeling with an emphasis on biological, biomedical, and behavioral sciences. Inspired by this work, this paper aims to focus on the benefits of ANN models, the most commonly used machine learning model in recent years, to the challenging problems of the design and analysis of composite materials and structures. Although ANN models are becoming popular topics in academia, the application of the ANN models has not been widely accepted as a standard module in the design and analysis of real composite materials and structures. 

This paper focuses on three important applications of ANN models in composite materials and structures: learning nonlinear constitutive model, accelerating multiscale modeling, and design optimization. 
It has been shown that the current nonlinear constitutive modeling is still not mature enough to accurately predict the nonlinear behaviors of composite materials \cite{hinton2004failure,hinton2013triaxial}. With the increasing complexity of the microstructures, the nonlinear behaviors of composite materials are becoming more and more complex, which often involves several failure mechanisms occurring simultaneously.
Moreover, some failure mechanisms cannot be fully captured by the existing physics-based models. In other words, the current physics-based constitutive models are not sufficient to accurately describe all the failure behaviors of composites. In addition to developing a constitutive model for a composite, many researchers employed a multiscale modeling approach to directly link the homogenized behavior at the macroscale to the corresponding microstructure, which does not require an explicit material model at the macroscale. A sub-scale modeling, either at microscale or mesoscale, is required for each integration point at the macroscale. Generally speaking, the sub-scale model will take the input such as deformation gradient and state variables from the macroscale model to perform dehomogenization so that the strain and stress fields at each material point at the sub-scale model can be obtained. Then, a nonlinear material model of each constituent (e.g., fiber and matrix for fiber-reinforced composites) will be used to degrade the material stiffness matrix based on the recovered local fields. A homogenization will be performed with the updated material stiffness to generate needed information (e.g., effective tangent stiffness matrix and macro-stress) for the macroscale analysis. Clearly, such expensive computations are not feasible for the practical engineering design and analysis \cite{geers2017homogenization}. Furthermore, multiscale modeling does not actually solve the problem of unknown physics-based models, but shift the problem to the sub-scale because the nonlinear materials models for the constituents and their interfaces are usually unknown. In addition to the above two issues, design optimization is another challenging and important topic in composite materials and structures. This challenge is also connected to the aforementioned two challenges. Without the accurate constitutive laws of constituents, the nonlinear behaviors of composite materials and structures cannot be well-captured. In addition, a key issue of the design optimization is to explore a huge design space in an efficient way, because there are often a number of design parameters across different scales and the corresponding physics-based models are very time-consuming.

The importance of incorporating data into the advanced materials design has been emphasized in Materials Genome Initiative (MGI) \cite{national2011materials}. One way to make good use of data is through machine learning techniques.
The machine learning used in the design and analysis of composites can be categorized by the training data -- whether the data is produced to inform the models or by the physical models. The data in the first scenario often comes from experiments. The machine learning models are used to describe phenomena in complex systems where we do not yet have a good physical understanding \cite{CMreport}. Learning unknown constitutive models is a typical application of this kind of machine learning models. The data in the second scenario usually comes from computer simulations where we have established physical models and machine learning is to provide an efficient tool to replace the expensive simulations and to interpret large scale computed data. Constructing surrogate models to accelerate multiscale modeling is a typical application of this kind of machine learning models. The ANN models are the most widely used machine learning models in recent years due to the outstanding performance with the growing data, ability to approximate complex nonlinear relations, and advanced open-source libraries (e.g., Tensorflow \cite{tensorflow2015-whitepaper} and PyTorch \cite{paszke2017automatic}), to name a few. As a result, ANN models are considered as an emerging technique in the design and analysis of composite materials and structures. To fully exploit the potentials of ANN models, there are at least two questions to be answered: what are the benefits from ANN models in design and analysis composites and how can we reap these benefits?

This paper is designed to provide the basic knowledge and state-of-the-art literature to point out the challenges and opportunities for those who are interested in learning and applying ANN models to composite materials and structures. Section 2 will briefly introduce the the basic algorithms of the feed-forward ANN model. Section 3 and 4 will discuss the usage of ANN models in learning unknown constitutive laws and accelerating multiscale modeling respectively. Section 5 will focus on the applications of ANN models in the design optimization of composites. Section 6 will conclude the review.

\section{Basic Algorithms of Feed-forward ANN Model}
\label{S:2}

\begin{figure}[!htb]
  \centering
  \includegraphics[width=0.6\textwidth]{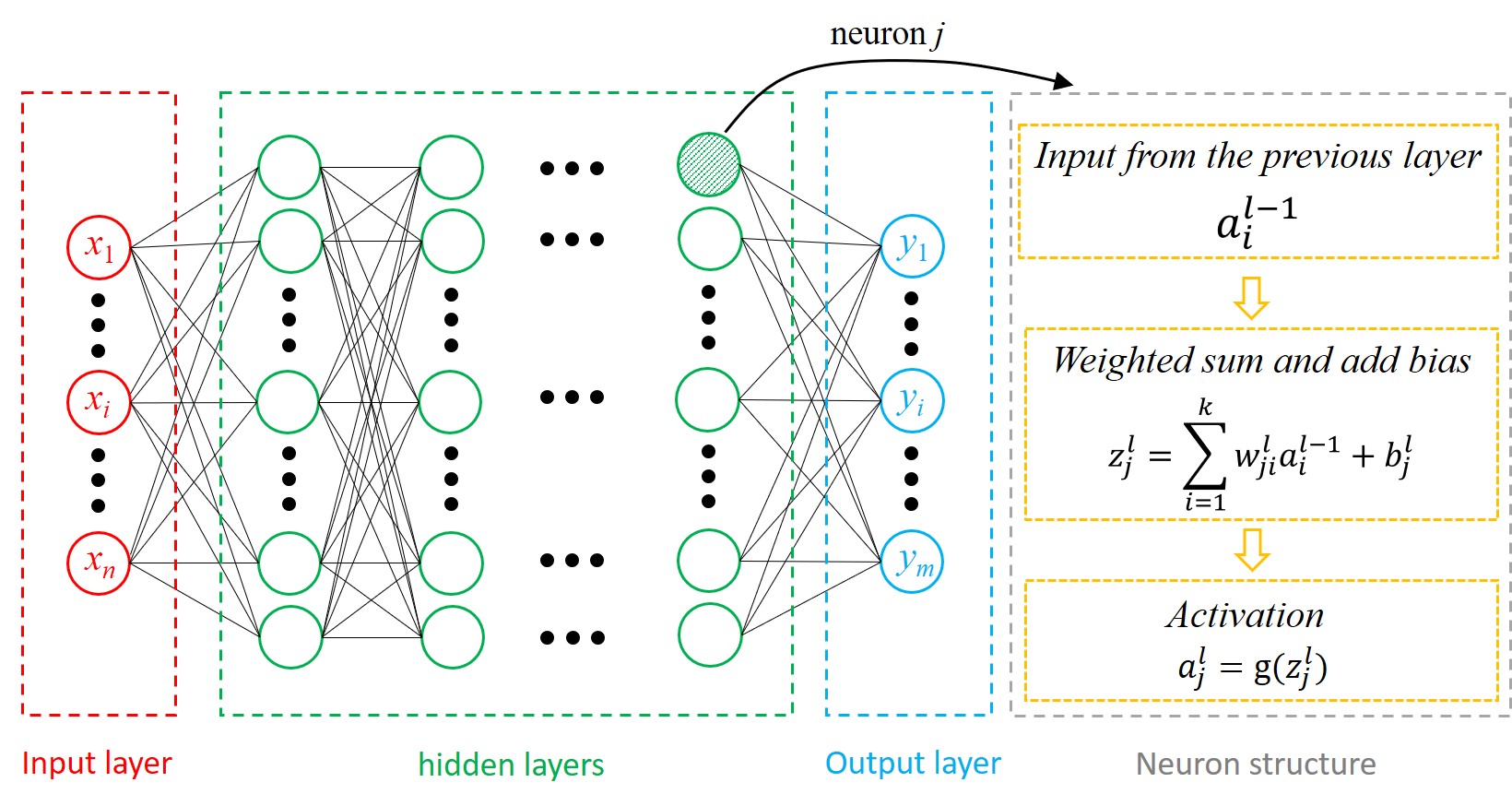}
  \caption{The architecture of a multilayer feed-forward network}
  \label{F:NN}
\end{figure}

The feed-forward ANN model, also called multilayer perception (MLP) model, is the most widely used ANN model and serves as the basis of other advanced ANN models such as the convolutional neural network (CNN) and recurrent neural network (RNN) \cite{goodfellow2016deep}. The architecture of a feed-forward ANN model is given in Fig. \ref{F:NN}, which contains three different layers: input layer, hidden layer(s), and output layer. The units in each layer are called neurons. Each neuron is connected to other neurons in the previous and next layer through weights. A weight parameter $w_{ji}^l$ measures the influence of the $i^{th}$ neuron in the $(l-1)^{th}$ layer on the $j^{th}$ neuron in the $l^{th}$ layer. A bias term $b_j^l$ is used in the $j^{th}$ neuron in the $l^{th}$ layer to cover a wider range \cite{nielsen2015neural}.

From Fig. \ref{F:NN}, the value of a neuron $a_j^l$ is obtained by adding the sum of the weighted neural inputs and a bias:
\begin{equation} \label{eq:zj}
  a_j^l = g \left(z_j^l\right) = g \left(\sum_{i=1}^k w_{ji}^l a_i^{l-1} + b_j^l \right)
\end{equation}
\noindent{w}here $k$ is the total number of neurons in the $\left( l-1 \right)^{th}$ layer. $g(\cdot)$ is defined as the activation function, which enables the ANN models to capture the nonlinear functions. In order to update the weights and biases in the model, a cost or objective function is required. For regression problems, the cost function is often defined as the mean square error (MSE) \cite{nielsen2015neural}:
\begin{equation} \label{eq:loss}
    L=\frac{1}{2n}\sum^{n}(y-\hat{y})^2+\text{regularization}
\end{equation}
where $n$ is the number of training samples, $y$ is the predicted value from the ANN model and $\hat{y}$ is the labeled output in the training dataset. The regularization term is used to prevent overfitting which has many different forms. A typical $L_2$ regularization term is defined as \cite{nielsen2015neural}:
\begin{equation} \label{eq:RLterm}
    L_2=\frac{\lambda}{2n}\sum w^2
\end{equation}
where $\lambda>0$ is the regularization parameter and $w$ is the weights of the neural networks. The weights and biases can be updated using the following equations \cite{haykin2010neural}:
\begin{equation}\label{eq:w}
    w_{ji}^{new} = w_{ji}^{old} - \eta \frac{\partial L}{\partial w_{ji}^{old}}
\end{equation}
\begin{equation}\label{eq:b}
    b_{j}^{new} = b_{j}^{old} - \eta \frac{\partial L}{\partial b_{j}^{old}}
\end{equation}
\noindent{w}here $\eta$ is the learning rate.

\section{Nonlinear Constitutive Models}
\label{S:3}

\subsection{Significance}

In general, there are two approaches to constructing a constitutive model of composites. The first approach is to develop or postulate a function (e.g., a polynomial) to approximate the homogenized material behaviors with several unknown parameters to be calibrated from experiments \cite{hahn1973nonlinear,hu2015failure}. The second approach is to build a constitutive model by homogenizing all the constitutive behaviors of constituents at the sub-scale. Note that the second approach also relies on the constitutive model of each constituent at the sub-scale, which still needs to be calibrated from experiments. 
Despite many developments in the two approaches, the nonlinear constitutive modeling of composite materials remains a critical challenge. With the increasing complexity of the microstructures in advanced composite materials, the nonlinear homogenized constitutive behavior at the macroscale is likely to be driven by one or multiple nonlinear mechanisms (e.g., viscosity, plasticity, or damage) at the sub-scale. To handle the increasing complexity, many ANN-based data-driven models have been developed. The basic idea is to capture the missing physics based on the experimental data and approximate the constitutive model with a function in a form-free manner. Since the aim of machine learning techniques is to identify the correlations in data and approximate such correlations using functions \cite{peng2020multiscale}, using machine learning to learn constitutive models has the potential to provide a good complement or even substitution to the physics-based models.

\subsection{Applications}

\begin{figure}[!htb]
  \centering
  \includegraphics[width=0.36\textwidth]{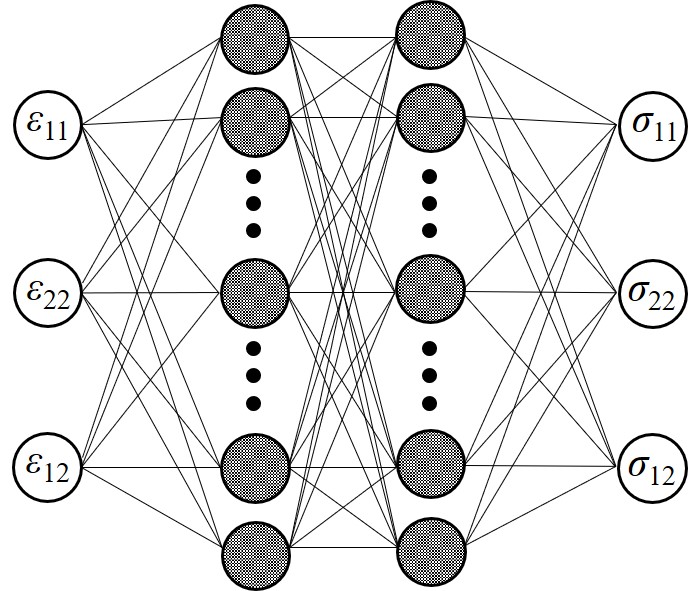}
  \caption{2D strain-stress relation constructed by an ANN model (gray circles are the hidden neurons)}
  \label{F:ANN}
\end{figure}

The traditional polynomial-based constitutive models are based on a function with coefficients to be determined from experiments. The issue is the inaccuracy caused by the assumptions associated with the postulated functions \cite{liu2019initial}. The problem is also called model errors, which remains inevitably as far as a material model is written explicitly \cite{furukawa1998implicit}. ANN models are widely used in approximating complex mappings between input and output as an universal approximation theorem \cite{csaji2001approximation}, which allows to approximate any functions in a form-free manner. In addition to the universal function approximation, the ANN model also features several other advantages in constructing a constitutive model \cite{xu2020learning} such as better performance for unevenly distributed data, approximating a non-smooth function, and mapping a high dimensional input-output. The early work of ANN-based constitutive models were carried out by Ghaboussi and his co-workers \cite{ghaboussi1991knowledge,ghaboussi1998new,ghaboussi1998autoprogressive,hashash2004numerical,jung2006neural,yun2008new}. The nonlinear material behaviors such as rate-dependent \cite{jung2006neural} and hysteretic behaviors \cite{yun2008new} were captured by the ANN models and the numerical implementation of the ANN models in finite element (FE) codes was also carried out \cite{hashash2004numerical}. Fig. \ref{F:ANN} shows a simple 2D strain-stress ANN model. Another important contribution of Ghaboussi's work is to enable learning the strain-stress relation by combining the data from structural responses with FE models \cite{ghaboussi1998autoprogressive}. The key motivation is to make use of experiments with rich constitutive information. The stress obtained from experiments are often derived from forces, which are not directly measurable. Further, it is very challenging to derive stress from the forces if the stress state is not uniform. For example, the stress in each layer of a multi-directional laminate in a 3-point bending test can hardly be derived. Therefore, for the ANN models based on the paired strain-stress data, the data can only come from simple experiments, where the homogenized stress can be derived from the measurable load-deflection curves. Clearly, the data from such simple experiments is not enough for training ANN models. By combining experimental data from structural tests and FE models, Ghaboussi and his co-workers developed the autoprogressive models to enlarge the available data from complex experiments by connecting load-deflection curves to strain-stress data. This idea was further extended to a self-learning FE method by several other researchers \cite{shin2000self,yun2008self}. Recently, Huang et al. \cite{huang2020learning} also developed ANN models for learning constitutive laws based on experimental data from structural tests and FE models. Different from the previous work, this method does not rely on directly paired strain-stress data, which is a critical requirement hindering ANN models from learning complex constitutive models. However, this method needs a full-field displacement measurement and is difficult to extend to three-dimensional (3D) problems and multiscale modeling problems. To enable ANN models to learn more complex constitutive models with limited measurements, Liu et al. \cite{liu2020learning,liu2020neural} and Xu et al. \cite{xu2020inverse} coupled ANN models into mechanical systems (e.g., FE models) so that the real input and output of ANN models are derived from the coupled system during the training. This idea has been applied to learn the nonlinear in-plane shear constitutive model \cite{liu2020neural}, failure initiation criterion \cite{liu2020neural}, damage accumulation law \cite{liu2020learning}, and viscoelastic material response \cite{xu2020inverse}.

In addition to ANN models, other machine learning models have also been developed to enable a model-free constitutive law. Kirchdoerfer and Ortiz \cite{kirchdoerfer2016data} developed data-driven computational mechanics to directly construct the constitutive law from the experimental material dataset. This method aims to find the point in the material dataset that is closest to satisfying the essential constraints and conservation laws. This approach has been further extended to solve inelasticity \cite{eggersmann2019model} and finite elasticity \cite{conti2020data} problems. Another machine learning technique for constructing the constitutive law is the manifold learning developed by Iba$\tilde{\text{n}}$ez and his co-workers, which has been applied to learn the elastic, inelastic, and plastic material models \cite{ibanez2017data,ibanez2018manifold,ibanez2019hybrid}. 

\subsection{Challenges and opportunities}

\subsubsection{Lack of Training Data}

Although some constitutive models were directly constructed based experimental data \cite{najjar2007simulating,mahdi2008crushing,zopf2017numerical,rodriguez2019application}, the constitutive relation is usually very simple such as one-dimensional (1D) strain-stress relation. However, the learned constitutive models will usually be used in a FE model for the structural analysis, which requires the 3D strain-stress relation. For a 3D constitutive model, there are at least 6 strain components as input and 6 stress components as output. If ANN models are used to learn more advanced material models such as path/history dependent material behavior, more input and output dimensions are required. To the best of our knowledge, there is no 3D constitutive laws constructed by ANN-based models with real experimental data. Clearly, it is extremely challenging to perform experiments to generate a representative training dataset, which includes 6-dimensional input and 6-dimensional output. However, there are a lot of data from complex experiments containing ample constitutive information. For example, a three-point bending test of a multi-directional laminate has different 3D strain-stress states at each layer, while only the load-deflection curve can be directly measured from experiments. Therefore, the key question is to derive or extract the rich constitutive information from experimental measurable data. 

\begin{figure}[!htb]
    \centering
    \begin{subfigure}{0.6\textwidth}
    \centering
        \includegraphics[width=\textwidth]{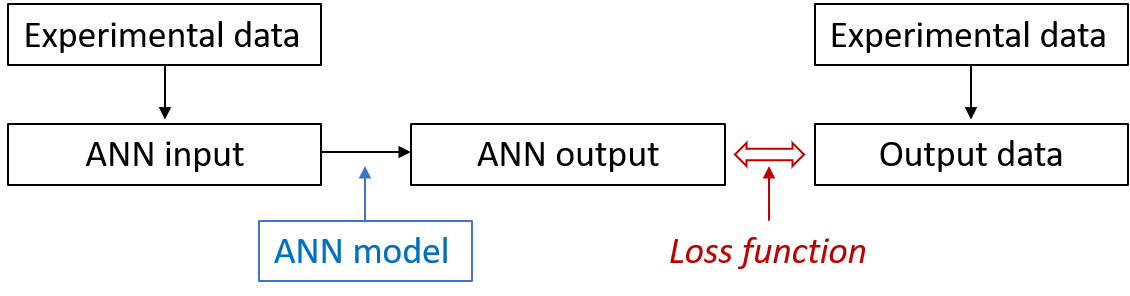}
        \caption{Training with directly paired data}
        \label{fig:dpd}
      \end{subfigure}
      \hfill
      \begin{subfigure}{0.85\textwidth}
      \centering
        \includegraphics[width=\textwidth]{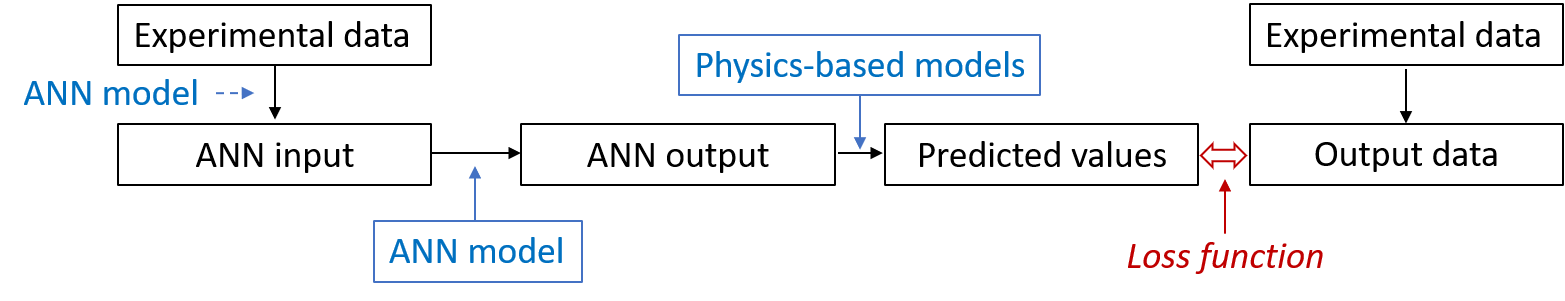}
        \caption{Training with indirectly measurable data}
        \label{fig:imd}
      \end{subfigure}
      \caption{Different methods of using training data}
      \label{F:data1}
    \end{figure}

Tang et al. \cite{tang2019map123} proposed a method called MAP123 that constructs a 3D constitutive model using just 1D data. The key idea is to expand the experimental measurable data (e.g., 1D data) to 3D strain-stress data. A finite deformation problem was analyzed by this method. Another potential solution is to couple ANN models into a mechanical model (e.g., a FE model) to derive the experimental measurable data. Fig. \ref{fig:dpd} shows the traditional training method that requires the directly paired data for training ANN models. Since the training data needs to be directly obtained or derived from experiments, it is hard to generate 3D strain-stress data for composite materials, because the output (3D stress) cannot be derived or measured from a load-deflection curve with an unknown constitutive law. Fig. \ref{fig:imd} shows a different method that the ANN output goes into other physics-based models to generate the predicted values which can be directly measured or derived from experiments. The loss function is constructed based on the predicted values and measurable data from experiments. Note that the predict values contain the trainable parameters (i.e., weights and biases) which are to be updated by minimizing the loss function. The benefit of the latter method is that the direct input and output data of the ANN model are not required to be measurable. For example, the ANN model still constructs strain and stress relation but using the load-deflection curve measured from experiments. 

\subsubsection{Coupling ANN with Mechanical Models}

The ANN coupled mechanical system posts another challenge that the input data may be derived from experimental data via the ANN model under training.
For example, the strain in a FE model is derived from displacements which are solved from the entire FE model with the ANN-based constitutive law. In other words, the input of the ANN model and the ANN model itself need to be determined simultaneously during the training \cite{liu2020learning,xu2020physics}. However, this approach brings another benefit that the physical constraints are implicitly imposed into the training process as the ANN model must go through the FE model to generate correct input data \cite{xu2020inverse,xu2020learning}. Some examples of using indirectly measurable data for training ANN models are given in Table \ref{tb:directdata}. 

\begin{table}[!htb]
  \centering
  \caption{Learn unknown physics based on indirectly measurable data}\label{tb:directdata}
  \begin{adjustbox}{width=0.98\textwidth}
  \begin{tabular}{|c|c|c|c|}
  \hline
  \textbf{Unknown physical laws} & \textbf{Mechanical model} & \textbf{Indirectly measurable data} & \textbf{ANN model} \\ \hline
  nonlinear in-plane shear behavior (lamina) \cite{liu2020neural}  & lamination theory & plate force ($F$), plate strain ($\epsilon$) & $f(\varepsilon_{12})=\sigma_{12}$ \\ \hline 
  failure criterion \cite{liu2020neural}  & lamination theory & failure load ($F$) & $f(\sigma_{ij})=\alpha$ \\ \hline 
  nonlinear in-plane shear behavior \cite{liu2020learning} & FE model & load ($F$), displacement ($u$) & $f(\varepsilon_{12})=\sigma_{12}$ \\ \hline 
  damage accumulation \cite{liu2020learning} & FE model + continuum damage model & macro-strain ($\bar{\varepsilon}_{ij}$), macro-stress ($\bar{\sigma}_{11}$) &  $f(\beta)=B$ \\ \hline 
  hyperelastic \cite{huang2020learning} & FE model & load ($F$), displacement ($u$) &  $f(\lambda)=P$ \\ \hline 
  elasto-plasticity \cite{xu2020learning} & FE model & load ($F$), displacement ($u$) &  $f(\varepsilon_{ij})=\sigma_{ij}$ \\ \hline 
  \end{tabular}
  \end{adjustbox}
  \end{table}

\subsubsection{Multiscale Inverse Modeling}

\begin{figure}[!htb]
  \centering
  \includegraphics[width=0.55\linewidth]{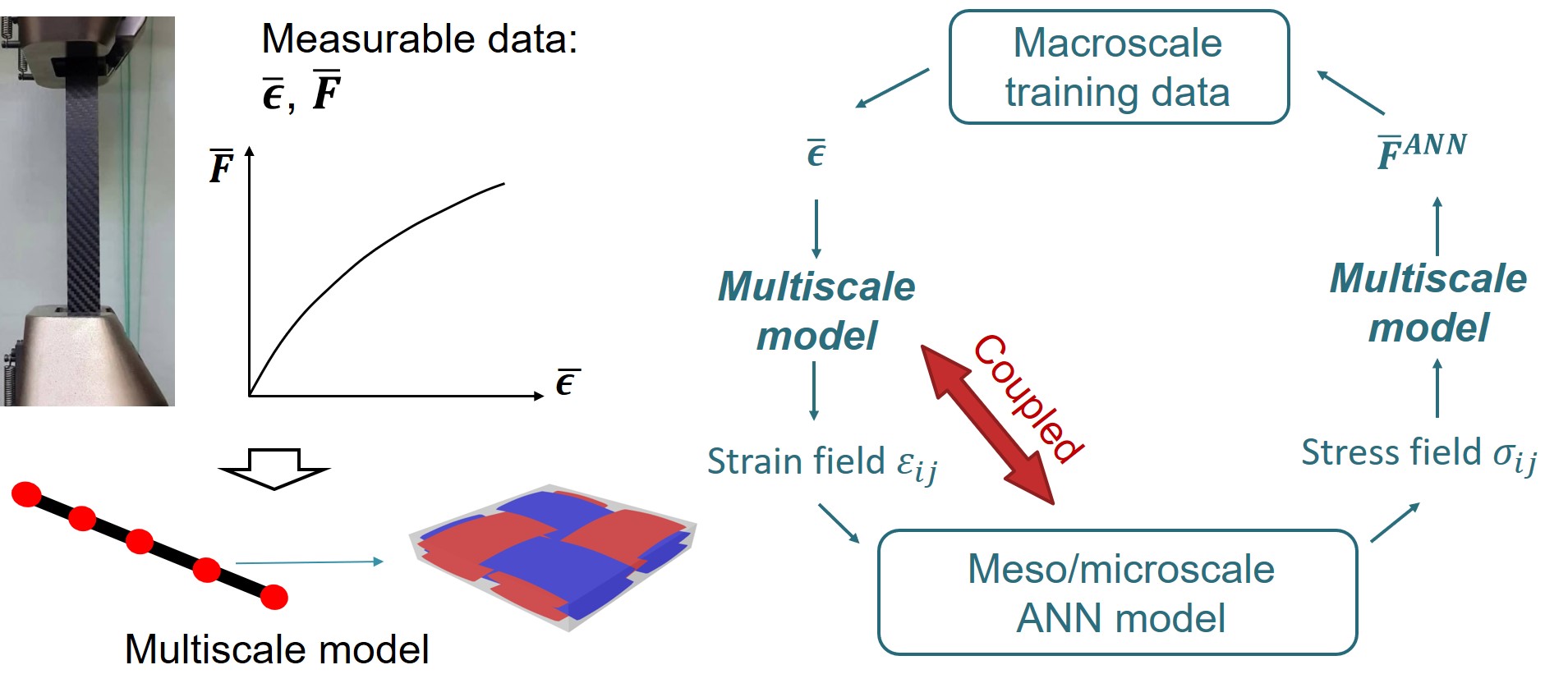}
  \caption{Train meso/microscale ANN model with measurable data from macroscale}
  \label{F:MLNNDL}
  \end{figure}

For composite materials and structures, there is a need to discover the constitutive laws of the constituents at the micro/mesoscale. However, it is hard to directly measure the properties of constituents (e.g., measuring transverse properties of fibers or 3D constitutive laws of fiber tows). The more common approach is to test a composite coupon and identify the properties of constituents using an inverse modeling, which is very challenging for nonlinear constitutive models. Some machine learning models have been used to help identify engineering constants at the sub-scale by a parameter inverse analysis \cite{bassir2009hybrid,komninelli2015towards}. However, identify the entire constitutive law at the sub-scale is a function inverse problems and the function space is infinite \cite{xu2020inverse}. The ANN coupled mechanical models could provide a solution which requires the ANN models to be coupled into a multiscale model (see Fig. \ref{F:MLNNDL}). In this way, the measurable information at the macroscale can be passed to the sub-scale to train the ANN models. However, effectively pass information across different scales and integrate ANN models within multiscale models are still challenging problems need to be solved.  


\section{Multiscale Modeling}
\label{S:4}

\subsection{Significance}

In the last decades, several advanced multiscale modeling approaches have been developed such as representative volume element (RVE) method based on finite element analysis (FEA) \cite{sun1996prediction,lomov2007meso}, mathematical homogenization theories (MHT) \cite{guedes1990preprocessing,fish1999computational}, mechanics of structure genome (MSG) \cite{yu2016unified,liu2017two,liu2019unified}, and generalized method of cells (GMC) and its extensions \cite{aboudi2004generalized,aboudi2012micromechanics}. These methods not only compute homogenized properties for the macroscale simulation (e.g., structural analysis) but also recover local stress and strain fields at the sub-scale via dehomogenization. For nonlinear homogenization problems, accurately recovering local stress and strain fields is crucial in determining the degradation of constituent properties, which subsequently changes the nonlinear macroscopic responses. Many commercial software tools have been developed based on the above methods (see Table \ref{tb:multiscale}) and some of them have been integrated into commercial FEA software to serve as a multiscale modeling module \cite{omairey2019development,liu2019multiscale}. Despite these advances, a key challenge of multiscale modeling in nonlinear structural analysis is the computational cost. 

\begin{table}[!htb]
\centering
\caption{Commercial multiscale modeling software}\label{tb:multiscale}
\begin{adjustbox}{width=0.7\textwidth}
\begin{tabular}{|c|c|c|}
\hline
Method & Software & Development team \\ \hline
FEA-based RVE & Digimat & e-Xstream Engineering \\ \hline 
FEA-based RVE & Material Designer & ANSYS \\ \hline 
FEA-based RVE & Micromechanics plugin & ABAQUS \\ \hline 
MHT & Multiscale Designer & Altair \\ \hline 
MSG & SwiftComp & AnalySwift \\ \hline 
GMC and its extensions & MAC/GMC & NASA Glenn Research Center \\ \hline 
\end{tabular}
\end{adjustbox}
\end{table}

In this paper, we will only deal with the continuum scales: microscale (constituent level), mesoscale (laminate/composite), and macroscale (global/structural level). At the macroscale level, the FE method is a standard tool for structural analysis. The homogenized properties deriving from sub-scale models will be used to define a structural element (e.g., beam, shell, or solid element) for the analysis. For nonlinear structural analysis, the homogenized properties are changing due to the nonlinear behaviors of constituents at the sub-scale. A popular modeling approach to capturing such nonlinear behavior is FE\textsuperscript{2} \cite{feyel2000fe2}. The macroscale FE-based structural analysis will use homogenized properties that are directly computed by the RVE analysis using the micro/mesoscale FE-based models, which means that there are two different FE-based models at macroscale (i.e., structural analysis) and micro/mesoscale (i.e., RVE analysis) respectively. Therefore, the FE\textsuperscript{2} removes the assumptions of the macroscale constitutive laws since there are no explicit material models used at the macroscale \cite{geers2010multi}. However, the computational effort can be extreme, even for a simple two-dimensional (2D) microstructure of fiber-reinforced composites. For advanced composite materials (e.g., textile composites or metamaterials), a 3D micro/mesostructure is needed. A total number of RVE calculations of a FE\textsuperscript{2} analysis can be approximated using the following equation \cite{yvonnet2019computational}:
\begin{equation}
  N_{tol} \approx \left(1+\beta\right) \times N_{int} \times N_{e} \times N_{iter} \times N_{evol}
\end{equation}
where $N_{int}$ is the number of integration points in a single element in the macroscopic mesh, $N_e$ is the total number of elements in the macroscopic mesh, $N_{iter}$ is the number of iterations in each time step, and $N_{evol}$ is the total time steps in a macroscopic analysis. Although in the literature using $\beta$=3 for 2D, however, it can only capture 2D behavior. To capture 3D behavior a 3D RVE is needed, thus $\beta$=6. For a real industrial problem, the nonlinear RVE analysis could be performed by millions of times, which makes FE\textsuperscript{2} impractical for real structural analyses.

\subsection{Applications}

\begin{figure}[!htb]
\centering
\includegraphics[width=0.85\linewidth]{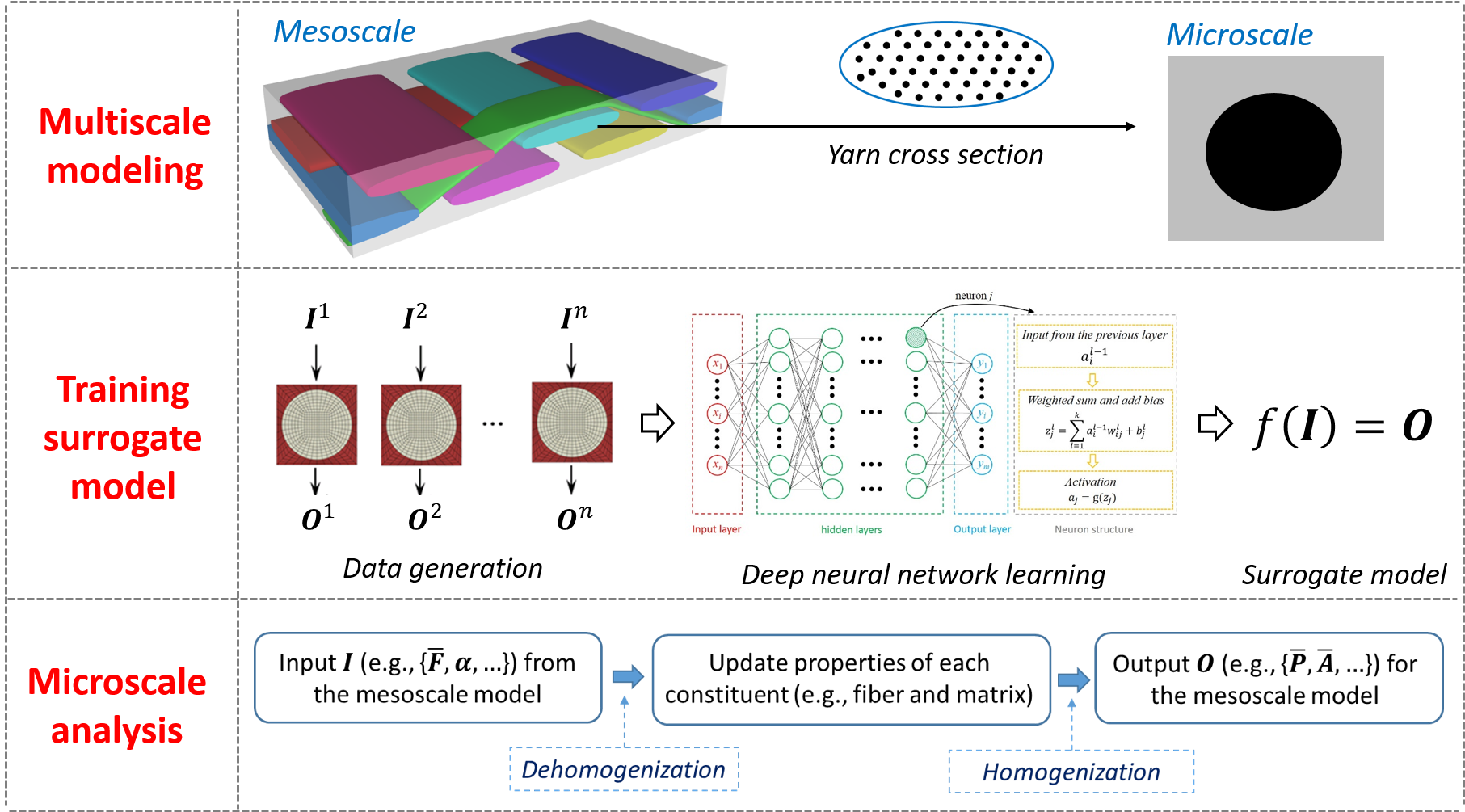}
\caption{General framework of data-driven multiscale modeling}
\label{F:multiscaleDNN}
\end{figure}

The whole multiscale modeling process can be approximated using a general function (Eq. (\ref{eq:IO})) with $\bm{I}$ as the input from the macroscale FE model and $\bm{O}$ as the output from the sub-scale models.
\begin{equation} \label{eq:IO}
    f\left( \bm{I} \right) = \bm{O}
\end{equation}
where $f$ can be the FE-based models (i.e., FE\textsuperscript{2}) or any functions describing the change of the state of a point in the macroscale model. The basic idea of using ANN models to accelerate multiscale modeling is to replace the FE-based model with a surrogate model constructed from data. A general framework of using a deep neural network model in nonlinear multiscale modeling is given in Fig. \ref{F:multiscaleDNN}. A 3D textile composite is used as an example. This deep neural network model is used to approximate the nonlinear behavior of a material point in a yarn of a textile composite, which is based on the data computed from a series of microscale analyses with fiber and matrix.

Lefik at al. \cite{lefik2003artificial, lefik2009artificial} applied ANN models in numerical modeling of composites. One of the applications in this work was to predict the homogenized nonlinear behavior (i.e. strain-stress curve) of a two-phase composite material. An elastic-plastic material model was constructed based on the data from a series of nonlinear microscale analyses. Unger and K$\ddot{\text{a}}$nke \cite{unger2009neural} applied an ANN model to construct a relation between macroscopic stress and crack opening responses. Le et al. \cite{le2015computational} performed a number of RVE analyses with periodic boundary conditions to generate training data for constructing a constitutive model for nonlinear elastic material behavior. The macroscopic strains as well as some microstructural parameters were used as input and the effective potential was computed as the output. The similar idea was applied to study the  nonlinear anisotropic electrical response of graphene/polymer nanocomposites \cite{lu2019data}. Liu and his co-works \cite{liu2016self,bessa2017framework,yan2018data,li2019clustering} developed various surrogate models to facilitate the multiscale modeling of composite materials (e.g., fiber-reinforced composite, amorphous material, and woven composite). They also incorporated the manufacturing process into the models. Sun and his co-workers \cite{wang2018multiscale,wang2019meta,wang2019cooperative,vlassis2020geometric} utilized advanced ANN models (e.g., RNN and deep reinforced learning (DRL)) to develop different surrogate models for multiscale modeling of multi-porosity materials. The cohesive laws were generated using a deep RNN model based on an offline homogenization procedure, which was proven to be much more computational efficient than FE\textsuperscript{2} approach \cite{wang2018multiscale}. The DRL was employed to construct surrogate models for the traction-separation law \cite{wang2019meta} and elasto-plasticity model \cite{wang2019cooperative}. In addition to the standard ANN models, Liu and his co-workers \cite{liu2019deep,liu2019exploring,liu2019transfer,liu2020deep} developed a deep material network for multiscale modeling of heterogeneous materials. This method has been used to construct surrogate models for history-dependent plasticity, finite strain hyperelasticity, and interfacial failure analysis. Recently, Liu et al. \cite{liu2020intelligent} extended this approach to develop an integrated framework for process modeling, material homogenization, machine learning, and multiscale simulation. Other researchers also applied different machine learning models to construct surrogate models to capture different nonlinear material behaviors such as elasto-plasticity \cite{wu2020bayesian,rocha2020micromechanics,mozaffar2019deep}, finite deformation hyperelasticity \cite{nguyen2020deep,yang2019derivation,sagiyama2019machine,fritzen2018two} and viscoplasticity \cite{yuan2018machine}. In addition to directly approximating multiscale modeling of composites, machine learning models have also been used to accelerate the expensive computations in FE models \cite{oishi2017computational,capuano2019smart}, which reduces the computing costs in many FE-based multiscale models. Note that ANN models are also employed in the homogenization of other physical behaviors such as electrical response \cite{lu2019data} and thermal conductivity \cite{wei2018predicting,rong2019predicting}. Table \ref{tb:ANNmodels} lists some homogenized constitutive behaviors approximated by different ANN models based on the data from simulations. Many ANN models have been developed for hyperelasticity constitutive behaviors due to the lower input and output dimensions (e.g., strains and elastic energy). Recently, much research interest has been focused on modeling the path/history-dependent behavior, which usually deals with high-dimensional input and output that may be better approximated by more advanced ANN models such as CNN, RNN or a mixed CNN-RNN model \cite{koeppe2020intelligent}.

\begin{table}[!htb]
  \centering
  \caption{Different nonlinear constitutive laws approximated by ANN models}\label{tb:ANNmodels}
  \begin{adjustbox}{width=0.7\textwidth}
  \begin{tabular}{|c|c|}
  \hline
  Nonlinear constitutive & ANN models  \\ \hline
  elastoplastic & feed-forward ANN \cite{lefik2003artificial, lefik2009artificial,yang2020exploring, zhang2020using}, RNN \cite{mozaffar2019deep,wu2020recurrent}, DRL \cite{wang2019meta} \\ \hline 
  viscoplastic  & feed-forward ANN \cite{stoffel2018artificial,stoffel2019neural}, RNN \cite{ghavamian2019accelerating}, CNN \cite{stoffel2020deep} \\ \hline 
  hyperelasticity & feed-forward ANN \cite{shen2005finite,le2015computational,sagiyama2019machine,nguyen2020deep,vlassis2020geometric} \\ \hline 
  damage & feed-forward ANN \cite{hambli2011multiscale, settgast2020hybrid} \\ \hline 
  traction-separation & feed-forward ANN \cite{fernandez2020application}, DRL \cite{wang2019meta} \\ \hline 
  \end{tabular}
  \end{adjustbox}
  \end{table}

\subsection{Challenges and opportunities}

\begin{figure}[!htb]
    \centering
    \includegraphics[width=0.9\linewidth]{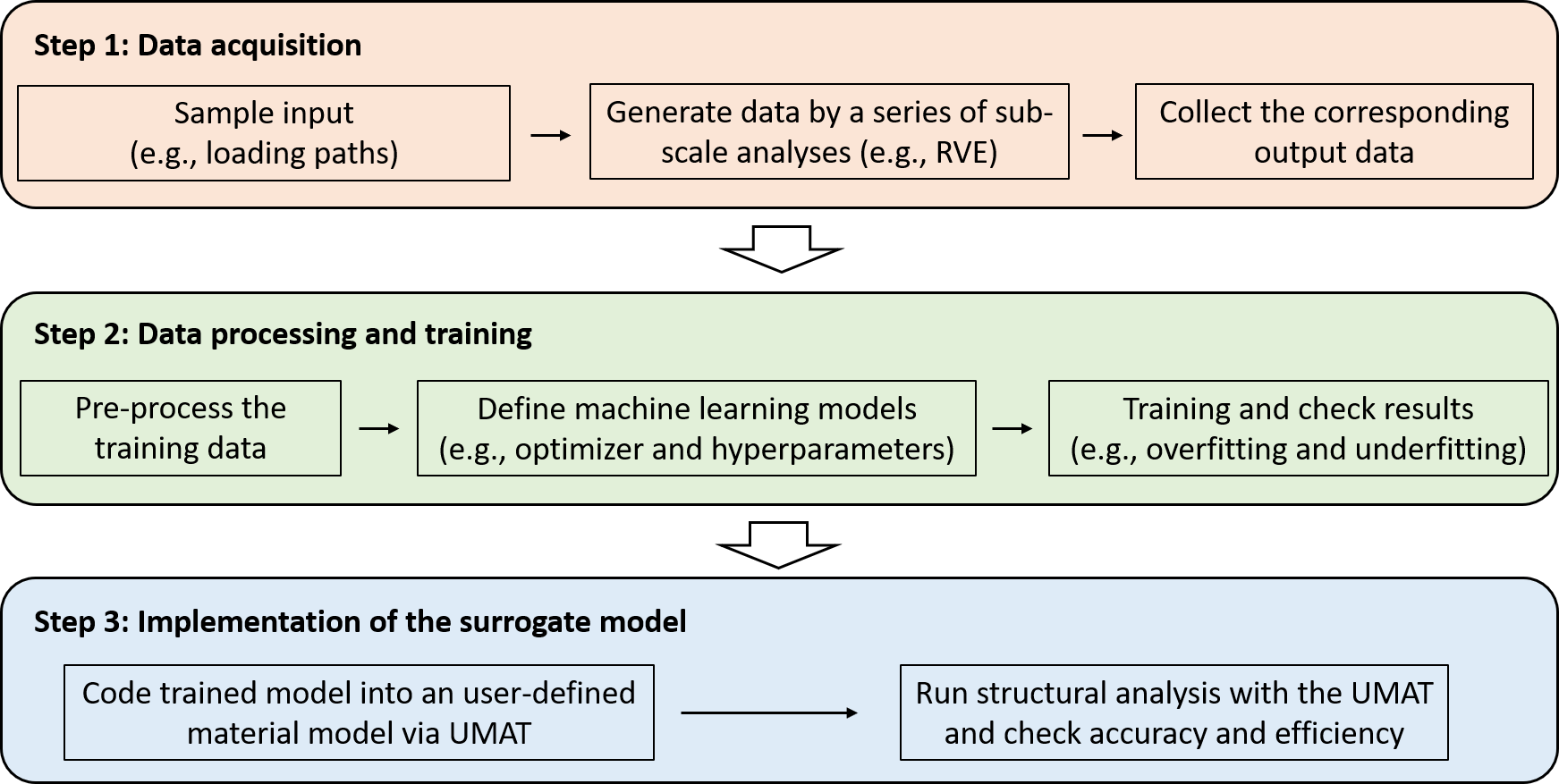}
    \caption{General procedure of multiscale surrogate modeling}
    \label{F:multisurrogate}
\end{figure}

Although various ANN models have been employed to construct surrogate models for different material responses, most of the works follow the same procedure. As shown in Fig. \ref{F:multisurrogate}, the first step is to generate the training dataset from a series of sub-scale analyses. The input could be the macroscopic loading paths, which are used to recover the local strain and stress fields at the sub-scale. Usually, several loading paths are required and a number of increments are needed in each loading path \cite{lefik2009artificial}. Once the input and output data are obtained from the first step, the second step is to train the surrogate model based on a machine learning model. A pre-processing of data (e.g., normalization) is often required in most machine learning models. Then, ANN models or other machine learning models are constructed aiming to develop a function like Eq. (\ref{eq:IO}) to approximate the relation between input and output data. The trained model needs to be tested to avoid overfitting or underfitting \cite{nielsen2015neural}. Once the desired surrogate model is trained, users need to code the model through a user-defined subroutine in FE software packages such as the UMAT in Abaqus to perform structural analysis \cite{systemes2014abaqus}. This surrogate model should provide a remarkable computational efficiency without losing too much accuracy compared with the high-fidelity multiscale analysis (e.g., FE\textsuperscript{2}). 

\subsubsection{Computational Cost of High-fidelity Simulation Models}

The challenge is the computational cost for generating training data. Simulation data from the high-fidelity multiscale analysis is very time-consuming. To construct nonlinear material models, a high-dimensional input and output data are usually required. For example, the path/history-dependent materials will require additional loading history or state variables as input. Some microstructural parameters (e.g., fiber volume fraction or microscale defects) maybe also needed as additional inputs to describe the microstructure of composites. Such high-dimensional input and output data make the machine learning model suffer the curse of dimensionality \cite{bessa2017framework}, which often requires the models to use more training data to get a good approximation. 

There are several potential solutions to reduce the size of the training data. Raissi et al. \cite{raissi2019physics} developed the physics-informed neural network (PINN) model to incorporate physical laws into the loss function (Eq. (\ref{eq:PINN})). 
\begin{equation} \label{eq:PINN}
    MSE = MSE_u + MSE_f
\end{equation}
where $MSE_u$ is the MSE from the training data and $MSE_f$ is the MSE from the imposed physical constraints. Fei et al. \cite{tao2020physics} showed that the size of training data can be reduced to only a fraction of the original one by adding some physical constraints for a thin shell buckling problem. Zhang et al.\cite{zhang2020PINN} incorporated physics knowledge into deep long short-term memory (LSTM) networks to study nonlinear structural responses subjected to a ground motion excitation. The results showed that the model with physical constraints can be trained with less rich data with a good accuracy. Following this idea, different physical laws have been developed and imposed to the PINN models such as the conservation law \cite{xu2020physics}, symmetric positive definite stiffness matrix \cite{xu2020learning}, and thermodynamics \cite{masi2020thermodynamics}. In addition to reducing the size of training data, appropriately adding physical constraints may also reduce the ill-posed optimization problems during the training \cite{xu2020learning}. 

Another solution is to keep developing more efficient yet accurate multiscale modeling approaches such as MSG, MHT and some reduced-order models (e.g., self-consistent clustering analysis \cite{liu2018microstructural}, proper orthogonal decomposition (POD) reduced model \cite{rocha2020adaptive}, and non-uniform transformation field analysis \cite{michel2010non}). 
Recently, the multi-fidelity modeling has been used to effectively generate training data for ANN models in composite materials and structures \cite{liu2019multi,yoo2020novel,tian2020multi}, which provides another approach to balance the accuracy and efficiency in training an ANN model \cite{meng2020composite}. Also, it would be attractive if a model can be tuned for both low- and high-fidelity models~\cite{wang2007review}.

\subsubsection{Physical Interpretation}

The ANN models also suffer the deficiency of the physical interpretation, which is a primary reason hindering machine learning models in the real engineering application as it is very hard for engineers to justify the results and for them to confidently use the model. Adding physical constraints (e.g., PINN) will enforce the trained ANN models to follow the known physical laws to some extent. Usually, the users need to adjust the ratio between the $MSE_u$ and $MSE_f$ in Eq. (\ref{eq:PINN}) so that the trained ANN model will result in a good prediction. In addition to PINN, physics-guided/theory-guided ANN models have been developed aiming to design the ANN architecture with the prior knowledge in physics \cite{yao2020fea,gao2020physics,zobeiry2020theory}. The FEA-Net has been recently proposed incorporating knowledge of FEA into deep convolutional neural network to estimate material properties of composite materials \cite{yao2020fea,gao2020physics}. Zobeiry at al. \cite{zobeiry2020theory} developed theory-guided ANN models by selecting features and ANN architectures with the knowledge of failure and strain-softening behavior in composites. Since the ANN models are designed with the prior knowledge in physics, the models are usually interpretable to the users. Furthermore, the theory-guided ANN models showed a good potential in the extrapolation. In other words, the models have a much improved prediction beyond the training zone, which cannot be handled well by conventional ANN models. However, the design of such ANN models are highly problem-dependent and largely driven by the domain knowledge of the users.

\section{Design Optimization}
\label{S:5}

\subsection{Significance}

One major advantage of advanced composite materials and structures is that their properties can be tailored for different requirements in different applications.
Materials can be placed and oriented in a way that their capabilities can be mostly utilized.
Besides, by carefully arrange constituent materials and design local structures, the complete part can achieve almost equally well performance for different applications without adding too much extra weights.
However, such a design process is challenging for this complex material-structural system mainly because of the expensive numerical simulation and large amount of design parameters available.

Hence, metamodels, including ANN, has been widely used in the past years and is receiving more attentions these days, for the purpose of reaching the global optimum in a fast, reliable and robust way.
Several papers on reviewing the use of metamodels in design optimization of composite materials and structures have been published in the past decade~\cite{simpson2004approximation,wang2007review,viana2014special,shu2017metamodel,jiang2020surrogate}.
Generally speaking, metamodels can help the design process in the following ways: 1) reducing the cost of computational analysis by approximating the expensive physical model, 2) providing overall knowledge of the problem in the design domain, 3) assisting engineers to define a better optimization model, and 4) supporting the optimization process.
Building an approximation (surrogate model) to the original physical model is the main direction for improving the optimization efficiency and main concern of this survey.

The benefit of cost-saving of utilizing surrogate models such as neural networks is manifold.
First, the surrogate model reduces the analysis time during each iteration of the optimization process.
The more analyses needed, the more significant this benefit will be.
For instance, though extremely popular, genetic algorithm \cite{muc2001genetic} is notoriously slow in convergence and requires large amount of functional evaluations.
On the other hand, gradient-based methods are generally inapplicable for black-box physical models.
However, for surrogate models, either finite difference or direct derivative can be computed.
Second, parallel computing techniques can be implemented.
Instead of evaluating expensive functions sequentially in the optimization process, the data acquisition step of building the surrogate model only need independent evaluations of different design points in the domain, which can be done concurrently.
Third, evaluations can be done fast and accurate in regions with dense points.
This becomes more significant for non-gradient methods as the optimization process going.
With the increasing of points in a local region, one should have more confidence of deducing the function value of a new point, and hence the evaluation of the physical model would be more like a waste of time.
Last but not least, preprocessing time can be saved.
Generating an accurate geometry from a limit number of design parameters and meshing the part with good quality can be time consuming and subject to flaws and fails of the preprocess.
Instead, surrogate models directly link design variables to the output, which can save the preprocessing cost significantly.

Though considering the optimization process only, many researchers conducted comparisons and provided numbers showing the reduction of cost.
Chen and Cheng~\cite{chen2010data} developed an ANN-based optimization strategy and at least 50\% of computational time reduction is observed.
D{\'\i}az et al.~\cite{diaz2016efficient} compared several surrogate models for the reliability-based design optimization of a stiffened panel. The author reported at least three orders of magnitude of improvements of running time when using ANN.
Truong et al.~\cite{truong2020artificial} also measured the time used by an ANN-based method for several applications, all of which are less than 10\% of the physical model.

\subsection{Application}

\begin{figure}[!htb]
  \centering
  \includegraphics[width=0.8\linewidth]{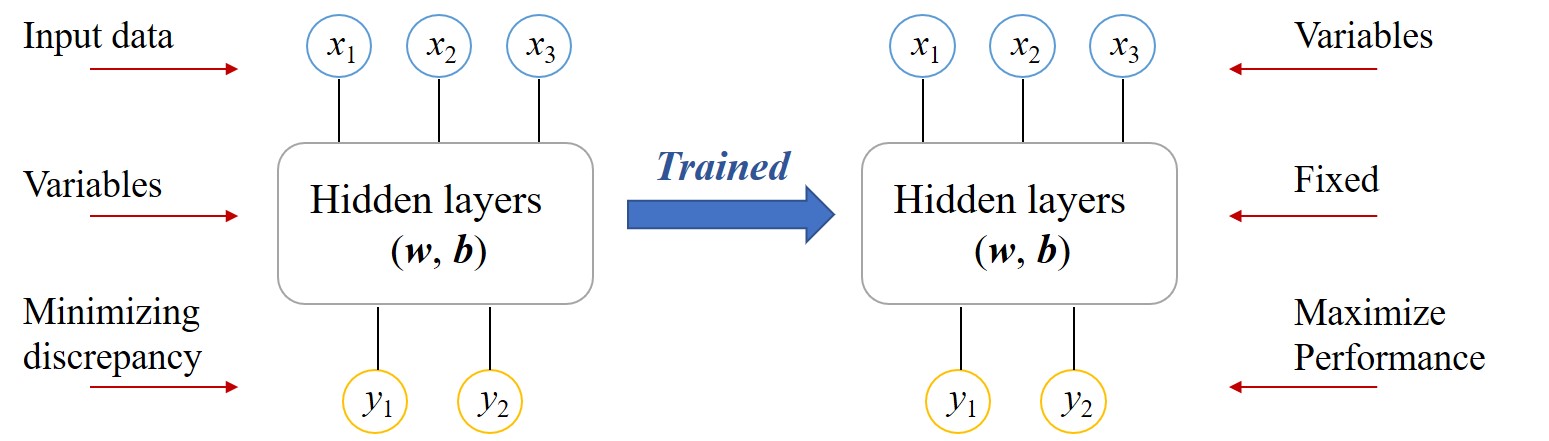}
  \caption{The general framework of machine learning assisted design optimization problems}
  \label{F:DO}
\end{figure}

The general framework using ANN models in design optimization problems is given in Fig. \ref{F:DO}. After training the model, the variables (i.e., weights $\bm{w}$ and biases $\bm{b}$) in the ANN model are determined. Then, the trained model is used to find the optimal inputs in the optimization problems. Such ANN models provide remarkable computational efficiency. First, similar to the aforementioned acceleration of multiscale modeling, ANN models are much efficient than the corresponding physics-based ones. Second, many physics-based design optimization methods rely on the commercial FE software which is used as a black-box in the analysis. It is difficult to apply the automatic differentiation or derive the analytical gradients of the model. Instead, finite difference method is often used in these analyses to calculate numerical gradients, which is very computationally expensive \cite{chen2020generative}. ANN models are differentiable that provide an efficient way to compute the gradient of the model. 

An early attempt of such implementations was performed to minimize the weight of the composite stiffened panels subjected to buckling constraints \cite{bisagni2002post}.
The key idea is to replace the expensive simulation (e.g., FE model) with ANN models.
Then, the commonly used optimization algorithms can be performed using the efficient ANN models.
The method was also used for a multi-objective optimization for vibration frequency and critical buckling load~\cite{abouhamze2007multi}.
The similar ideas were further developed for optimizing the weight and the failure load of composite stiffened panels \cite{marin2012optimization}.
Though one advantage of using ANN models is the potential of good approximations of the relation between high-dimensional input and output, most studies at this stage have no more than 10 design variables.
Typical inputs include the geometries of structural features such as stiffeners~\cite{bisagni2002post,marin2012optimization,fu2015minimum,yazdi2016optimization}, layup schemes (e.g. numbers of plies, material orientations and stacking sequence)~\cite{bisagni2002post,abouhamze2007multi,garmsiri2014multiobjective,gomes2017design}, and the architecture of the micro/mesoscale structures~\cite{fu2015minimum}.

In addition to using feed-forward ANN models in the shape and size optimization of composite structures, the application of CNN models to topology optimization problems has generated considerable interests recently, especially in the design of architectured materials thanks to the advances in additive manufacturing \cite{abueidda2020topology,wilt2020accelerating}.
A widely-used topology optimization strategy is to find the best constituent for each element in a fixed meshed domain, which is commonly described as a `pixel' or `voxel' style.
Because of this fundamental different feature from the shape and size optimization, and also the similarity to the task of image recognition, CNN is considered suitable for this purpose.
The CNN models take the input such as structure geometry, external loads, and boundary conditions and predict the stress fields \cite{nie2020stress}, structural stiffness \cite{banga20183d} or damage initiation \cite{li2019clustering}.
The trained model is then used to replace the physics-based models to perform the topology optimization.
In addition, CNN models are also used in topology optimization of composite materials by arranging stiff and soft materials to improve the fracture toughness \cite{gu2018novo,chen2020generative}.

Besides the independent tasks of the aforementioned optimization, a truly automatic and integrated optimization process (including topology, shape and size) is highly expected by the industry.
Because of the difference between the descriptions of structures, there is a gap between the finish of topology optimization and the start of shape and size optimization, where the geometry needs to be interpreted from `pixel art' and re-described using feature parameters.
Yildiz et al.~\cite{yildiz2003integrated} used an ANN model with a single hidden layer for recognizing different types of holes from the `pixel' result of the topology optimization.
In this case, the neural network is not the approximation of a physical model anymore.
Instead, it serves as a bridge connecting different optimization stages and a support to the broader design process.

\subsection{Challenges and opportunities}

\subsubsection{Computational Cost in Sampling}

The improvement of efficiency acquired from using ANN models becomes not so significant when taking the sampling cost into account.
This comes with two aspects: the number (or dimension) of design variables and the size of the design space, both of which affect the total number of sampling points needed.
There are several approaches developed to reduced the data size.
Data-analysis has been used by some researchers to tackle this challenge~\cite{chen2010data,wang2020data}.
For the design optimization problems, the design variables are usually not equally important to the objective performance.
Therefore, the dimension of the input can be reduced by performing a sensitivity analysis to remove the negligible design variables \cite{xiang2014prediction,balokas2018neural}.
Moreover, some data-processing approach can also be used to reduce the sampling cost \cite{wang2020data,han2020efficient}.
Besides, classification or clustering analysis can be used to recognize feasible/infeasible regions in the design space~\cite{chen2010data}.
Hence, if a future sampling point is located in the infeasible region without any doubt, there is no need to run the point through the simulation of objectives, and computational resources can be reserved for more points in the feasible region.

A large-scale problem is challenging because of the sampling method and strategy.
Since the physical model is expensive, knowing where to draw samples is valuable.
The classical methods of design and analysis of computer experiments (DACE) such as central composite design or Box-Behnken design are inapplicable because of their exponential growth with the dimension of the design space.
The most popular method in the literature is the Latin hypercube sampling (LHS), since the number of samples needed has weak relations to the design space dimension and mainly depends on the number of divisions along each dimension.
There are drawbacks, however, that the divisions must be the same for all dimensions and the sample size must be doubled each time when more data is needed.
For sampling strategy, one-shot sampling is adopted by most studies, which naturally requires a statistically uniform distribution of samples in the whole design space.
Recent years have seen a growing interest on adaptive sampling for drawing samples intelligently and achieving a better distribution of samples and hence less evaluations of physical models.
However, there are not many studies utilizing this strategy in the community of ANN-based design optimization of composite structures.


\subsubsection{Uncertainties in Design Optimization}

Incorporating uncertainties in the design of composite materials and structures is another critical and challenging problem.
There are uncertainties associated with composition, constituent properties, and defects which strongly influence the performance of composite materials and structures \cite{UQreport}.
The uncertainties quantification is also a compute-intensive problem where ANN models can be used as an efficient alternative to physics-based models \cite{balokas2018neural,acar2020machine,bessa2017framework}.
Two major and popular topics under the broader concept of optimization under uncertainty (OUU) are reliability-based design optimization (RBDO) and robust design optimization (RDO).
The former one concerns the probability of constraints satisfaction because of the randomness of variables, while the latter one tries to reduce the effects from random variables on the change of objectives.
For the ANN-based composites design community, though RBDO has received much attention in recent years~\cite{gomes2011reliability,diaz2016efficient,song2017multi,yoo2020novel}, research on RDO is still in its very early stage.
Besides, regarding to the source of uncertainties, most studies focused on the design variables, and little has been done for quantifying uncertainties of the ANN model itself.

\section{Conclusions}

Although ANN models have been used to solve various problems in composite materials and structures, most of them are developed for only two purposes, namely accelerating simulations and discovering unknown physics. The general procedure of accelerating simulations using ANN models can be summarized as follows:

\begin{enumerate}
  \item Sample the input data of the model. The sampling should be designed to ensure the trained ANN model to have a good predictive performance with the new input data. In the meanwhile, to reduce the computational cost, users need to minimize the size of dataset by reducing the dimensions of input features and the number of samples.
  \item Perform computer simulations. In general, a number of physics-based simulations are performed to generate the paired input and output data.
  \item Train the ANN model. Apply the techniques in training ANN models such as selecting the architecture of the ANN model, tuning the hyperparameters, and cross validation, to name a few.
  \item Replace the expensive simulation with the trained ANN models. For a multiscale modeling problem, the ANN model usually needs to be coded into commercial FE software through an user-defined subroutine. 
\end{enumerate}
Note that accelerating modeling of composite materials and structures are not restricted to multiscale constitutive modeling. The ANN models have been applied to many other problems in the modeling of composite materials and structures such as predicting ballistic limit \cite{artero2018influence}, first-ply failure \cite{ang2018first}, and the effects of embedded defects and features \cite{el2018multiscale}. The major challenge is to reduce the cost of data generation from physics-based models and add interpretability to the ANN model. Several research directions are discussed in the paper such as adding physical-constraints, developing more efficient physics-based models, and using multi-fidelity data.

Discovering unknown physics is essentially an inverse problem. Learning complex 2D or 3D nonlinear constitutive laws directly from experiments requires a huge number of training data, which is usually not feasible for the experiments of composite materials and structures. A potential solution is to use indirect data derived from experimental measurements through some physics-based models, which will bring several benefits. First, the data from complex experiments, containing rich constitutive information, becomes available as the directly paired data is not required. Second, the real ANN output needs to go through another physics-based model, which enforces certain physical constraints in the loss function. Third, this approach enables learning nonlinear constitutive laws at the constituents scale with the data from coupon or structural tests. For example, this approach can be extended to learn nonlinear constitutive laws of interface between fiber and matrix using the structural responses of the composite. However, this research is still in its early stages and the key challenge is to connect the data from ANN models and physics-based models across different scales.

\begin{figure}[!htb]
  \centering
  \includegraphics[width=0.5\linewidth]{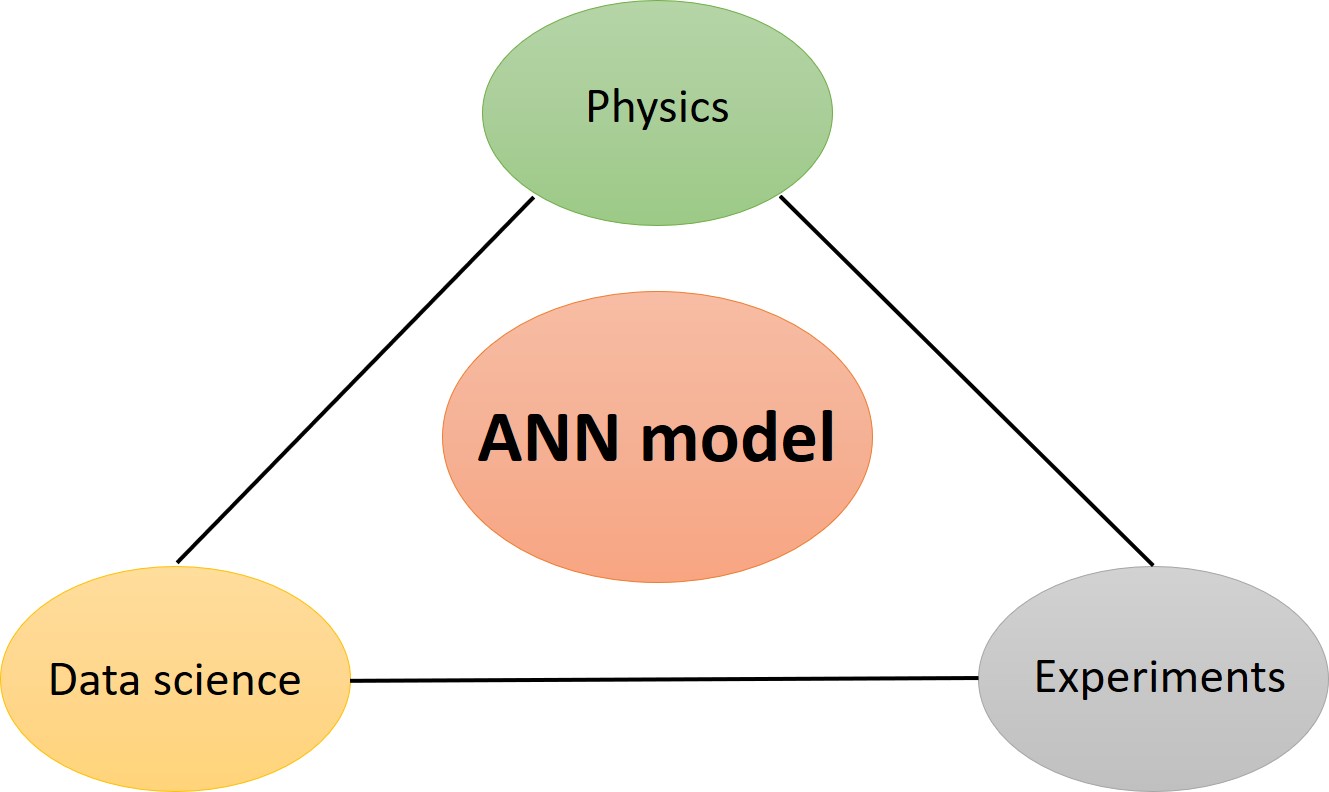}
  \caption{Integrated ANN model with physics, experiments, and data science}
  \label{F:tri}
\end{figure}

In short, the recent efforts in the modeling of composite materials and structures shows a trend towards applying ANN models in conjunction with physics, experiments and data science, which creates a hybrid system utilizing full potentials from each element (see Fig. \ref{F:tri}). 
Bring physics into ANN models will reduce the requirement of the size of training datasets, reduce the potential physically inconsistent prediction \cite{karpatne2017physics}, and enhance the physical interpretability. Appropriately design experiments with enriched constitutive information combined with physics-based models will expand the data availability for training ANN models. The methods from data science can be used to pre-process the training data (e.g., principal component analysis \cite{yang2020prediction}) to reduce the input dimensions which will greatly reduce the computational costs.
The key challenge is to connect different elements into a single system so that the ANN models can keep tracking the information across different element.

We hope that this review provides an introduction and overview of ANN models in the design and analysis of composite materials and structures, which will stimulate discussion and highlight some future research directions in the composites community.

\bibliography{sample}

\newcommand{\noop}[1]{}
\begin{thebibliography}{151}
\newcommand{\enquote}[1]{``#1''}
\providecommand{\natexlab}[1]{#1}
\providecommand{\url}[1]{\texttt{#1}}
\providecommand{\urlprefix}{URL }
\expandafter\ifx\csname urlstyle\endcsname\relax
  \providecommand{\doi}[1]{doi:\discretionary{}{}{}#1}\else
  \providecommand{\doi}{doi:\discretionary{}{}{}\begingroup
  \urlstyle{rm}\Url}\fi

\bibitem[{Dimiduk et~al.(2018)Dimiduk, Holm, and
  Niezgoda}]{dimiduk2018perspectives}
Dimiduk, D.~M., Holm, E.~A., and Niezgoda, S.~R., \enquote{Perspectives on the
  impact of machine learning, deep learning, and artificial intelligence on
  materials, processes, and structures engineering,} \emph{Integrating
  Materials and Manufacturing Innovation}, Vol.~7, No.~3, 2018, pp. 157--172.

\bibitem[{Peng et~al.(2020)Peng, Alber, Tepole, Cannon, De, Dura-Bernal,
  Garikipati, Karniadakis, Lytton, Perdikaris et~al.}]{peng2020multiscale}
Peng, G.~C., Alber, M., Tepole, A.~B., Cannon, W.~R., De, S., Dura-Bernal, S.,
  Garikipati, K., Karniadakis, G., Lytton, W.~W., Perdikaris, P., et~al.,
  \enquote{Multiscale Modeling Meets Machine Learning: What Can We Learn?}
  \emph{Archives of Computational Methods in Engineering}, 2020, pp. 1--21.

\bibitem[{Hinton et~al.(2004)Hinton, Soden, and Kaddour}]{hinton2004failure}
Hinton, M., Soden, P., and Kaddour, A.-S., \emph{Failure Criteria in Fibre
  Reinforced Polymer Composites: The World-wide Failure Exercise}, Elsevier,
  2004.

\bibitem[{Hinton and Kaddour(2013)}]{hinton2013triaxial}
Hinton, M., and Kaddour, A., \enquote{Triaxial test results for
  fibre-reinforced composites: The Second World-Wide Failure Exercise benchmark
  data,} \emph{Journal of Composite Materials}, Vol.~47, No. 6-7, 2013, pp.
  653--678.

\bibitem[{Geers et~al.(2017)Geers, Kouznetsova, Matou{\v{s}}, and
  Yvonnet}]{geers2017homogenization}
Geers, M.~G., Kouznetsova, V.~G., Matou{\v{s}}, K., and Yvonnet, J.,
  \enquote{Homogenization methods and multiscale modeling: nonlinear problems,}
  \emph{Encyclopedia of Computational Mechanics Second Edition}, 2017, pp.
  1--34.

\bibitem[{Science and (US)(2011)}]{national2011materials}
Science, N., and (US), T.~C., \emph{Materials genome initiative for global
  competitiveness}, Executive Office of the President, National Science and
  Technology Council, 2011.

\bibitem[{Dolbow et~al.(2019)Dolbow, Garikipati, and Park}]{CMreport}
Dolbow, J.~E., Garikipati, K., and Park, H.~S., \enquote{Computational
  Mechanics Vision and Future Challenges Workshop Report,} Tech. rep.,
  University of Michigan, Ann Arbor, MI, 2019.

\bibitem[{Abadi et~al.(2015)Abadi, Agarwal, Barham, Brevdo, Chen, Citro,
  Corrado, Davis, Dean, Devin, Ghemawat, Goodfellow, Harp, Irving, Isard, Jia,
  Jozefowicz, Kaiser, Kudlur, Levenberg, Man\'{e}, Monga, Moore, Murray, Olah,
  Schuster, Shlens, Steiner, Sutskever, Talwar, Tucker, Vanhoucke, Vasudevan,
  Vi\'{e}gas, Vinyals, Warden, Wattenberg, Wicke, Yu, and
  Zheng}]{tensorflow2015-whitepaper}
Abadi, M., Agarwal, A., Barham, P., Brevdo, E., Chen, Z., Citro, C., Corrado,
  G.~S., Davis, A., Dean, J., Devin, M., Ghemawat, S., Goodfellow, I., Harp,
  A., Irving, G., Isard, M., Jia, Y., Jozefowicz, R., Kaiser, L., Kudlur, M.,
  Levenberg, J., Man\'{e}, D., Monga, R., Moore, S., Murray, D., Olah, C.,
  Schuster, M., Shlens, J., Steiner, B., Sutskever, I., Talwar, K., Tucker, P.,
  Vanhoucke, V., Vasudevan, V., Vi\'{e}gas, F., Vinyals, O., Warden, P.,
  Wattenberg, M., Wicke, M., Yu, Y., and Zheng, X., \enquote{{TensorFlow}:
  Large-Scale Machine Learning on Heterogeneous Systems,} , 2015.
\newblock \urlprefix\url{https://www.tensorflow.org/}, software available from
  tensorflow.org.

\bibitem[{Paszke et~al.(2017)Paszke, Gross, Chintala, Chanan, Yang, DeVito,
  Lin, Desmaison, Antiga, and Lerer}]{paszke2017automatic}
Paszke, A., Gross, S., Chintala, S., Chanan, G., Yang, E., DeVito, Z., Lin, Z.,
  Desmaison, A., Antiga, L., and Lerer, A., \enquote{Automatic differentiation
  in PyTorch,} \emph{In NIPS 2017 Autodiff Workshop: The Future of
  Gradient-based Machine Learning Software and Techniques}, 2017.

\bibitem[{Goodfellow et~al.(2016)Goodfellow, Bengio, Courville, and
  Bengio}]{goodfellow2016deep}
Goodfellow, I., Bengio, Y., Courville, A., and Bengio, Y., \emph{Deep
  learning}, Vol.~1, MIT press Cambridge, 2016.

\bibitem[{Nielsen(2015)}]{nielsen2015neural}
Nielsen, M.~A., \emph{Neural networks and deep learning}, Vol. 2018,
  Determination press San Francisco, CA, USA:, 2015.

\bibitem[{Haykin(2010)}]{haykin2010neural}
Haykin, S., \emph{Neural Networks and Learning Machines, 3/E}, Pearson
  Education India, 2010.

\bibitem[{Hahn and Tsai(1973)}]{hahn1973nonlinear}
Hahn, H.~T., and Tsai, S.~W., \enquote{Nonlinear elastic behavior of
  unidirectional composite laminae,} \emph{Journal of Composite Materials},
  Vol.~7, No.~1, 1973, pp. 102--118.

\bibitem[{Hu et~al.(2015)Hu, Lin, and Tu}]{hu2015failure}
Hu, H.-T., Lin, W.-P., and Tu, F.-T., \enquote{Failure analysis of
  fiber-reinforced composite laminates subjected to biaxial loads,}
  \emph{Composites Part B: Engineering}, Vol.~83, 2015, pp. 153--165.

\bibitem[{Liu et~al.(2019{\natexlab{a}})Liu, Gasco, Goodsell, and
  Yu}]{liu2019initial}
Liu, X., Gasco, F., Goodsell, J., and Yu, W., \enquote{Initial failure strength
  prediction of woven composites using a new yarn failure criterion constructed
  by deep learning,} \emph{Composite Structures}, Vol. 230, 2019{\natexlab{a}},
  p. 111505.

\bibitem[{Furukawa and Yagawa(1998)}]{furukawa1998implicit}
Furukawa, T., and Yagawa, G., \enquote{Implicit constitutive modelling for
  viscoplasticity using neural networks,} \emph{International Journal for
  Numerical Methods in Engineering}, Vol.~43, No.~2, 1998, pp. 195--219.

\bibitem[{Cs{\'a}ji(2001)}]{csaji2001approximation}
Cs{\'a}ji, B.~C., \enquote{Approximation with artificial neural networks,}
  \emph{Faculty of Sciences, Etvs Lornd University, Hungary}, Vol.~24, 2001,
  p.~48.

\bibitem[{Xu et~al.(2020{\natexlab{a}})Xu, Huang, and Darve}]{xu2020learning}
Xu, K., Huang, D.~Z., and Darve, E., \enquote{Learning constitutive relations
  using symmetric positive definite neural networks,} \emph{arXiv preprint
  arXiv:2004.00265}, 2020{\natexlab{a}}.

\bibitem[{Ghaboussi et~al.(1991)Ghaboussi, Garrett~Jr, and
  Wu}]{ghaboussi1991knowledge}
Ghaboussi, J., Garrett~Jr, J., and Wu, X., \enquote{Knowledge-based modeling of
  material behavior with neural networks,} \emph{Journal of Engineering
  Mechanics}, Vol. 117, No.~1, 1991, pp. 132--153.

\bibitem[{Ghaboussi and Sidarta(1998)}]{ghaboussi1998new}
Ghaboussi, J., and Sidarta, D., \enquote{New nested adaptive neural networks
  (NANN) for constitutive modeling,} \emph{Computers and Geotechnics}, Vol.~22,
  No.~1, 1998, pp. 29--52.

\bibitem[{Ghaboussi et~al.(1998)Ghaboussi, Pecknold, Zhang, and
  Haj-Ali}]{ghaboussi1998autoprogressive}
Ghaboussi, J., Pecknold, D.~A., Zhang, M., and Haj-Ali, R.~M.,
  \enquote{Autoprogressive training of neural network constitutive models,}
  \emph{International Journal for Numerical Methods in Engineering}, Vol.~42,
  No.~1, 1998, pp. 105--126.

\bibitem[{Hashash et~al.(2004)Hashash, Jung, and
  Ghaboussi}]{hashash2004numerical}
Hashash, Y., Jung, S., and Ghaboussi, J., \enquote{Numerical implementation of
  a neural network based material model in finite element analysis,}
  \emph{International Journal for Numerical Methods in Engineering}, Vol.~59,
  No.~7, 2004, pp. 989--1005.

\bibitem[{Jung and Ghaboussi(2006)}]{jung2006neural}
Jung, S., and Ghaboussi, J., \enquote{Neural network constitutive model for
  rate-dependent materials,} \emph{Computers \& Structures}, Vol.~84, No.
  15-16, 2006, pp. 955--963.

\bibitem[{Yun et~al.(2008{\natexlab{a}})Yun, Ghaboussi, and
  Elnashai}]{yun2008new}
Yun, G.~J., Ghaboussi, J., and Elnashai, A.~S., \enquote{A new neural
  network-based model for hysteretic behavior of materials,}
  \emph{International Journal for Numerical Methods in Engineering}, Vol.~73,
  No.~4, 2008{\natexlab{a}}, pp. 447--469.

\bibitem[{Shin and Pande(2000)}]{shin2000self}
Shin, H., and Pande, G., \enquote{On self-learning finite element codes based
  on monitored response of structures,} \emph{Computers and Geotechnics},
  Vol.~27, No.~3, 2000, pp. 161--178.

\bibitem[{Yun et~al.(2008{\natexlab{b}})Yun, Ghaboussi, and
  Elnashai}]{yun2008self}
Yun, G.~J., Ghaboussi, J., and Elnashai, A.~S., \enquote{Self-learning
  simulation method for inverse nonlinear modeling of cyclic behavior of
  connections,} \emph{Computer Methods in Applied Mechanics and Engineering},
  Vol. 197, No. 33-40, 2008{\natexlab{b}}, pp. 2836--2857.

\bibitem[{Huang et~al.(2020)Huang, Xu, Farhat, and Darve}]{huang2020learning}
Huang, D.~Z., Xu, K., Farhat, C., and Darve, E., \enquote{Learning constitutive
  relations from indirect observations using deep neural networks,}
  \emph{Journal of Computational Physics}, Vol. 416, 2020, p. 109491.

\bibitem[{Liu et~al.(2020{\natexlab{a}})Liu, Tao, Du, Yu, and
  Xu}]{liu2020learning}
Liu, X., Tao, F., Du, H., Yu, W., and Xu, K., \enquote{Learning nonlinear
  constitutive laws using neural network models based on indirectly measurable
  data,} \emph{Journal of Applied Mechanics}, Vol.~87, No.~8,
  2020{\natexlab{a}}.

\bibitem[{Liu et~al.(2020{\natexlab{b}})Liu, Tao, and Yu}]{liu2020neural}
Liu, X., Tao, F., and Yu, W., \enquote{A neural network enhanced system for
  learning nonlinear constitutive law and failure initiation criterion of
  composites using indirectly measurable data,} \emph{Composite Structures},
  Vol. 252, 2020{\natexlab{b}}, p. 112658.

\bibitem[{Xu et~al.(2020{\natexlab{b}})Xu, Tartakovsky, Burghardt, and
  Darve}]{xu2020inverse}
Xu, K., Tartakovsky, A.~M., Burghardt, J., and Darve, E., \enquote{Inverse
  Modeling of Viscoelasticity Materials using Physics Constrained Learning,}
  \emph{arXiv preprint arXiv:2005.04384}, 2020{\natexlab{b}}.

\bibitem[{Kirchdoerfer and Ortiz(2016)}]{kirchdoerfer2016data}
Kirchdoerfer, T., and Ortiz, M., \enquote{Data-driven computational mechanics,}
  \emph{Computer Methods in Applied Mechanics and Engineering}, Vol. 304, 2016,
  pp. 81--101.

\bibitem[{Eggersmann et~al.(2019)Eggersmann, Kirchdoerfer, Reese, Stainier, and
  Ortiz}]{eggersmann2019model}
Eggersmann, R., Kirchdoerfer, T., Reese, S., Stainier, L., and Ortiz, M.,
  \enquote{Model-free data-driven inelasticity,} \emph{Computer Methods in
  Applied Mechanics and Engineering}, Vol. 350, 2019, pp. 81--99.

\bibitem[{Conti et~al.(2020)Conti, M{\"u}ller, and Ortiz}]{conti2020data}
Conti, S., M{\"u}ller, S., and Ortiz, M., \enquote{Data-Driven Finite
  Elasticity,} \emph{Archive for Rational Mechanics and Analysis}, Vol. 237,
  2020, pp. 1--33.

\bibitem[{Iba{\~n}ez et~al.(2017)Iba{\~n}ez, Borzacchiello, Aguado,
  Abisset-Chavanne, Cueto, Ladev{\`e}ze, and Chinesta}]{ibanez2017data}
Iba{\~n}ez, R., Borzacchiello, D., Aguado, J.~V., Abisset-Chavanne, E., Cueto,
  E., Ladev{\`e}ze, P., and Chinesta, F., \enquote{Data-driven non-linear
  elasticity: constitutive manifold construction and problem discretization,}
  \emph{Computational Mechanics}, Vol.~60, No.~5, 2017, pp. 813--826.

\bibitem[{Ibanez et~al.(2018)Ibanez, Abisset-Chavanne, Aguado, Gonzalez, Cueto,
  and Chinesta}]{ibanez2018manifold}
Ibanez, R., Abisset-Chavanne, E., Aguado, J.~V., Gonzalez, D., Cueto, E., and
  Chinesta, F., \enquote{A manifold learning approach to data-driven
  computational elasticity and inelasticity,} \emph{Archives of Computational
  Methods in Engineering}, Vol.~25, No.~1, 2018, pp. 47--57.

\bibitem[{Ib{\'a}{\~n}ez et~al.(2019)Ib{\'a}{\~n}ez, Abisset-Chavanne,
  Gonz{\'a}lez, Duval, Cueto, and Chinesta}]{ibanez2019hybrid}
Ib{\'a}{\~n}ez, R., Abisset-Chavanne, E., Gonz{\'a}lez, D., Duval, J.-L.,
  Cueto, E., and Chinesta, F., \enquote{Hybrid constitutive modeling:
  data-driven learning of corrections to plasticity models,}
  \emph{International Journal of Material Forming}, Vol.~12, No.~4, 2019, pp.
  717--725.

\bibitem[{Najjar and Huang(2007)}]{najjar2007simulating}
Najjar, Y.~M., and Huang, C., \enquote{Simulating the stress--strain behavior
  of Georgia kaolin via recurrent neuronet approach,} \emph{Computers and
  Geotechnics}, Vol.~34, No.~5, 2007, pp. 346--361.

\bibitem[{Mahdi and El~Kadi(2008)}]{mahdi2008crushing}
Mahdi, E.-S., and El~Kadi, H., \enquote{Crushing behavior of laterally
  compressed composite elliptical tubes: experiments and predictions using
  artificial neural networks,} \emph{Composite Structures}, Vol.~83, No.~4,
  2008, pp. 399--412.

\bibitem[{Zopf and Kaliske(2017)}]{zopf2017numerical}
Zopf, C., and Kaliske, M., \enquote{Numerical characterisation of uncured
  elastomers by a neural network based approach,} \emph{Computers \&
  Structures}, Vol. 182, 2017, pp. 504--525.

\bibitem[{Rodr{\'\i}guez-S{\'a}nchez et~al.(2019)Rodr{\'\i}guez-S{\'a}nchez,
  Ledesma-Orozco, Ledesma, and Vidal-Lesso}]{rodriguez2019application}
Rodr{\'\i}guez-S{\'a}nchez, A.~E., Ledesma-Orozco, E., Ledesma, S., and
  Vidal-Lesso, A., \enquote{Application of artificial neural networks to map
  the mechanical response of a thermoplastic elastomer,} \emph{Materials
  Research Express}, Vol.~6, No.~7, 2019, p. 075320.

\bibitem[{Tang et~al.(2019)Tang, Zhang, Yang, Li, Liu, and
  Guo}]{tang2019map123}
Tang, S., Zhang, G., Yang, H., Li, Y., Liu, W.~K., and Guo, X.,
  \enquote{MAP123: A data-driven approach to use 1D data for 3D nonlinear
  elastic materials modeling,} \emph{Computer Methods in Applied Mechanics and
  Engineering}, Vol. 357, 2019, p. 112587.

\bibitem[{Xu and Darve(2020)}]{xu2020physics}
Xu, K., and Darve, E., \enquote{Physics constrained learning for data-driven
  inverse modeling from sparse observations,} \emph{arXiv preprint
  arXiv:2002.10521}, 2020.

\bibitem[{Bassir et~al.(2009)Bassir, Guessasma, and
  Boubakar}]{bassir2009hybrid}
Bassir, D.~H., Guessasma, S., and Boubakar, L., \enquote{Hybrid computational
  strategy based on ANN and GAPS: application for identification of a
  non-linear model of composite material,} \emph{Composite Structures},
  Vol.~88, No.~2, 2009, pp. 262--270.

\bibitem[{Komninelli et~al.(2015)Komninelli, Iliopoulos, and
  Michopoulos}]{komninelli2015towards}
Komninelli, F., Iliopoulos, A., and Michopoulos, J.~G., \enquote{Towards
  identification of lower scale composite material properties from higher scale
  experimental data via inverse analysis of coupled multiscale models,}
  \emph{ASME 2015 International Design Engineering Technical Conferences and
  Computers and Information in Engineering Conference}, American Society of
  Mechanical Engineers Digital Collection, 2015.

\bibitem[{Sun and Vaidya(1996)}]{sun1996prediction}
Sun, C., and Vaidya, R., \enquote{Prediction of composite properties from a
  representative volume element,} \emph{Composites Science and Technology},
  Vol.~56, No.~2, 1996, pp. 171--179.

\bibitem[{Lomov et~al.(2007)Lomov, Ivanov, Verpoest, Zako, Kurashiki, Nakai,
  and Hirosawa}]{lomov2007meso}
Lomov, S.~V., Ivanov, D.~S., Verpoest, I., Zako, M., Kurashiki, T., Nakai, H.,
  and Hirosawa, S., \enquote{Meso-FE modelling of textile composites: Road map,
  data flow and algorithms,} \emph{Composites Science and Technology}, Vol.~67,
  No.~9, 2007, pp. 1870--1891.

\bibitem[{Guedes and Kikuchi(1990)}]{guedes1990preprocessing}
Guedes, J., and Kikuchi, N., \enquote{Preprocessing and postprocessing for
  materials based on the homogenization method with adaptive finite element
  methods,} \emph{Computer Methods in Applied Mechanics and Engineering},
  Vol.~83, No.~2, 1990, pp. 143--198.

\bibitem[{Fish et~al.(1999)Fish, Yu, and Shek}]{fish1999computational}
Fish, J., Yu, Q., and Shek, K., \enquote{Computational damage mechanics for
  composite materials based on mathematical homogenization,}
  \emph{International Journal for Numerical Methods in Engineering}, Vol.~45,
  No.~11, 1999, pp. 1657--1679.

\bibitem[{Yu(2016)}]{yu2016unified}
Yu, W., \enquote{A unified theory for constitutive modeling of composites,}
  \emph{Journal of Mechanics of Materials and Structures}, Vol.~11, No.~4,
  2016, pp. 379--411.

\bibitem[{Liu et~al.(2017)Liu, Rouf, Peng, and Yu}]{liu2017two}
Liu, X., Rouf, K., Peng, B., and Yu, W., \enquote{Two-step homogenization of
  textile composites using mechanics of structure genome,} \emph{Composite
  Structures}, Vol. 171, 2017, pp. 252--262.

\bibitem[{Liu et~al.(2019{\natexlab{b}})Liu, Yu, Gasco, and
  Goodsell}]{liu2019unified}
Liu, X., Yu, W., Gasco, F., and Goodsell, J., \enquote{A unified approach for
  thermoelastic constitutive modeling of composite structures,}
  \emph{Composites Part B: Engineering}, Vol. 172, 2019{\natexlab{b}}, pp.
  649--659.

\bibitem[{Aboudi(2004)}]{aboudi2004generalized}
Aboudi, J., \enquote{The generalized method of cells and high-fidelity
  generalized method of cells micromechanical models—A review,}
  \emph{Mechanics of Advanced Materials and Structures}, Vol.~11, No. 4-5,
  2004, pp. 329--366.

\bibitem[{Aboudi et~al.(2012)Aboudi, Arnold, and
  Bednarcyk}]{aboudi2012micromechanics}
Aboudi, J., Arnold, S.~M., and Bednarcyk, B.~A., \emph{Micromechanics of
  composite materials: a generalized multiscale analysis approach},
  Butterworth-Heinemann, 2012.

\bibitem[{Omairey et~al.(2019)Omairey, Dunning, and
  Sriramula}]{omairey2019development}
Omairey, S.~L., Dunning, P.~D., and Sriramula, S., \enquote{Development of an
  ABAQUS plugin tool for periodic RVE homogenisation,} \emph{Engineering with
  Computers}, Vol.~35, No.~2, 2019, pp. 567--577.

\bibitem[{Liu et~al.(2019{\natexlab{c}})Liu, Gasco, Yu, Goodsell, and
  Rouf}]{liu2019multiscale}
Liu, X., Gasco, F., Yu, W., Goodsell, J., and Rouf, K., \enquote{Multiscale
  analysis of woven composite structures in MSC.Nastran,} \emph{Advances in
  Engineering Software}, Vol. 135, 2019{\natexlab{c}}, p. 102677.

\bibitem[{Feyel and Chaboche(2000)}]{feyel2000fe2}
Feyel, F., and Chaboche, J.-L., \enquote{FE2 multiscale approach for modelling
  the elastoviscoplastic behaviour of long fibre SiC/Ti composite materials,}
  \emph{Computer Methods in Applied Mechanics and Engineering}, Vol. 183, No.
  3-4, 2000, pp. 309--330.

\bibitem[{Geers et~al.(2010)Geers, Kouznetsova, and
  Brekelmans}]{geers2010multi}
Geers, M.~G., Kouznetsova, V.~G., and Brekelmans, W., \enquote{Multi-scale
  computational homogenization: Trends and challenges,} \emph{Journal of
  Computational and Applied Mathematics}, Vol. 234, No.~7, 2010, pp.
  2175--2182.

\bibitem[{Yvonnet(2019)}]{yvonnet2019computational}
Yvonnet, J., \emph{Computational Homogenization of Heterogeneous Materials with
  Finite Elements}, Springer, 2019.

\bibitem[{Lefik and Schrefler(2003)}]{lefik2003artificial}
Lefik, M., and Schrefler, B.~A., \enquote{Artificial neural network as an
  incremental non-linear constitutive model for a finite element code,}
  \emph{Computer Methods in Applied Mechanics and Engineering}, Vol. 192, No.
  28-30, 2003, pp. 3265--3283.

\bibitem[{Lefik et~al.(2009)Lefik, Boso, and Schrefler}]{lefik2009artificial}
Lefik, M., Boso, D., and Schrefler, B., \enquote{Artificial neural networks in
  numerical modelling of composites,} \emph{Computer Methods in Applied
  Mechanics and Engineering}, Vol. 198, No. 21-26, 2009, pp. 1785--1804.

\bibitem[{Unger and K{\"o}nke(2009)}]{unger2009neural}
Unger, J.~F., and K{\"o}nke, C., \enquote{Neural networks as material models
  within a multiscale approach,} \emph{Computers \& Structures}, Vol.~87, No.
  19-20, 2009, pp. 1177--1186.

\bibitem[{Le et~al.(2015)Le, Yvonnet, and He}]{le2015computational}
Le, B., Yvonnet, J., and He, Q.-C., \enquote{Computational homogenization of
  nonlinear elastic materials using neural networks,} \emph{International
  Journal for Numerical Methods in Engineering}, Vol. 104, No.~12, 2015, pp.
  1061--1084.

\bibitem[{Lu et~al.(2019)Lu, Giovanis, Yvonnet, Papadopoulos, Detrez, and
  Bai}]{lu2019data}
Lu, X., Giovanis, D.~G., Yvonnet, J., Papadopoulos, V., Detrez, F., and Bai,
  J., \enquote{A data-driven computational homogenization method based on
  neural networks for the nonlinear anisotropic electrical response of
  graphene/polymer nanocomposites,} \emph{Computational Mechanics}, Vol.~64,
  No.~2, 2019, pp. 307--321.

\bibitem[{Liu et~al.(2016)Liu, Bessa, and Liu}]{liu2016self}
Liu, Z., Bessa, M., and Liu, W.~K., \enquote{Self-consistent clustering
  analysis: an efficient multi-scale scheme for inelastic heterogeneous
  materials,} \emph{Computer Methods in Applied Mechanics and Engineering},
  Vol. 306, 2016, pp. 319--341.

\bibitem[{Bessa et~al.(2017)Bessa, Bostanabad, Liu, Hu, Apley, Brinson, Chen,
  and Liu}]{bessa2017framework}
Bessa, M., Bostanabad, R., Liu, Z., Hu, A., Apley, D.~W., Brinson, C., Chen,
  W., and Liu, W.~K., \enquote{A framework for data-driven analysis of
  materials under uncertainty: Countering the curse of dimensionality,}
  \emph{Computer Methods in Applied Mechanics and Engineering}, Vol. 320, 2017,
  pp. 633--667.

\bibitem[{Yan et~al.(2018)Yan, Lin, Kafka, Lian, Yu, Liu, Yan, Wolff, Wu,
  Ndip-Agbor et~al.}]{yan2018data}
Yan, W., Lin, S., Kafka, O.~L., Lian, Y., Yu, C., Liu, Z., Yan, J., Wolff, S.,
  Wu, H., Ndip-Agbor, E., et~al., \enquote{Data-driven multi-scale
  multi-physics models to derive process--structure--property relationships for
  additive manufacturing,} \emph{Computational Mechanics}, Vol.~61, No.~5,
  2018, pp. 521--541.

\bibitem[{Li et~al.(2019)Li, Kafka, Gao, Yu, Nie, Zhang, Tajdari, Tang, Guo, Li
  et~al.}]{li2019clustering}
Li, H., Kafka, O.~L., Gao, J., Yu, C., Nie, Y., Zhang, L., Tajdari, M., Tang,
  S., Guo, X., Li, G., et~al., \enquote{Clustering discretization methods for
  generation of material performance databases in machine learning and design
  optimization,} \emph{Computational Mechanics}, Vol.~64, No.~2, 2019, pp.
  281--305.

\bibitem[{Wang and Sun(2018)}]{wang2018multiscale}
Wang, K., and Sun, W., \enquote{A multiscale multi-permeability poroplasticity
  model linked by recursive homogenizations and deep learning,} \emph{Computer
  Methods in Applied Mechanics and Engineering}, Vol. 334, 2018, pp. 337--380.

\bibitem[{Wang and Sun(2019)}]{wang2019meta}
Wang, K., and Sun, W., \enquote{Meta-modeling game for deriving
  theory-consistent, microstructure-based traction--separation laws via deep
  reinforcement learning,} \emph{Computer Methods in Applied Mechanics and
  Engineering}, Vol. 346, 2019, pp. 216--241.

\bibitem[{Wang et~al.(2019)Wang, Sun, and Du}]{wang2019cooperative}
Wang, K., Sun, W., and Du, Q., \enquote{A cooperative game for automated
  learning of elasto-plasticity knowledge graphs and models with AI-guided
  experimentation,} \emph{Computational Mechanics}, Vol.~64, No.~2, 2019, pp.
  467--499.

\bibitem[{Vlassis et~al.(2020)Vlassis, Ma, and Sun}]{vlassis2020geometric}
Vlassis, N., Ma, R., and Sun, W., \enquote{Geometric deep learning for
  computational mechanics Part I: Anisotropic Hyperelasticity,} \emph{arXiv
  preprint arXiv:2001.04292}, 2020.

\bibitem[{Liu et~al.(2019{\natexlab{d}})Liu, Wu, and Koishi}]{liu2019deep}
Liu, Z., Wu, C., and Koishi, M., \enquote{A deep material network for
  multiscale topology learning and accelerated nonlinear modeling of
  heterogeneous materials,} \emph{Computer Methods in Applied Mechanics and
  Engineering}, Vol. 345, 2019{\natexlab{d}}, pp. 1138--1168.

\bibitem[{Liu and Wu(2019)}]{liu2019exploring}
Liu, Z., and Wu, C., \enquote{Exploring the 3d architectures of deep material
  network in data-driven multiscale mechanics,} \emph{Journal of the Mechanics
  and Physics of Solids}, Vol. 127, 2019, pp. 20--46.

\bibitem[{Liu et~al.(2019{\natexlab{e}})Liu, Wu, and Koishi}]{liu2019transfer}
Liu, Z., Wu, C., and Koishi, M., \enquote{Transfer learning of deep material
  network for seamless structure--property predictions,} \emph{Computational
  Mechanics}, Vol.~64, No.~2, 2019{\natexlab{e}}, pp. 451--465.

\bibitem[{Liu(2020)}]{liu2020deep}
Liu, Z., \enquote{Deep material network with cohesive layers: Multi-stage
  training and interfacial failure analysis,} \emph{Computer Methods in Applied
  Mechanics and Engineering}, Vol. 363, 2020, p. 112913.

\bibitem[{Liu et~al.(2020{\natexlab{c}})Liu, Wei, Huang, and
  Wu}]{liu2020intelligent}
Liu, Z., Wei, H., Huang, T., and Wu, C., \enquote{Intelligent multiscale
  simulation based on process-guided composite database,} \emph{arXiv preprint
  arXiv:2003.09491}, 2020{\natexlab{c}}.

\bibitem[{Wu et~al.(2020{\natexlab{a}})Wu, Zulueta, Major, Arriaga, and
  Noels}]{wu2020bayesian}
Wu, L., Zulueta, K., Major, Z., Arriaga, A., and Noels, L., \enquote{Bayesian
  inference of non-linear multiscale model parameters accelerated by a Deep
  Neural Network,} \emph{Computer Methods in Applied Mechanics and
  Engineering}, Vol. 360, 2020{\natexlab{a}}, p. 112693.

\bibitem[{Rocha et~al.(2020{\natexlab{a}})Rocha, Kerfriden, and van~der
  Meer}]{rocha2020micromechanics}
Rocha, I., Kerfriden, P., and van~der Meer, F., \enquote{Micromechanics-based
  surrogate models for the response of composites: A critical comparison
  between a classical mesoscale constitutive model, hyper-reduction and neural
  networks,} \emph{European Journal of Mechanics-A/Solids}, Vol.~82,
  2020{\natexlab{a}}, p. 103995.

\bibitem[{Mozaffar et~al.(2019)Mozaffar, Bostanabad, Chen, Ehmann, Cao, and
  Bessa}]{mozaffar2019deep}
Mozaffar, M., Bostanabad, R., Chen, W., Ehmann, K., Cao, J., and Bessa, M.,
  \enquote{Deep learning predicts path-dependent plasticity,} \emph{Proceedings
  of the National Academy of Sciences}, Vol. 116, No.~52, 2019, pp.
  26414--26420.

\bibitem[{Nguyen-Thanh et~al.(2020)Nguyen-Thanh, Zhuang, and
  Rabczuk}]{nguyen2020deep}
Nguyen-Thanh, V.~M., Zhuang, X., and Rabczuk, T., \enquote{A deep energy method
  for finite deformation hyperelasticity,} \emph{European Journal of
  Mechanics-A/Solids}, Vol.~80, 2020, p. 103874.

\bibitem[{Yang et~al.(2019)Yang, Guo, Tang, and Liu}]{yang2019derivation}
Yang, H., Guo, X., Tang, S., and Liu, W.~K., \enquote{Derivation of
  heterogeneous material laws via data-driven principal component expansions,}
  \emph{Computational Mechanics}, Vol.~64, No.~2, 2019, pp. 365--379.

\bibitem[{Sagiyama and Garikipati(2019)}]{sagiyama2019machine}
Sagiyama, K., and Garikipati, K., \enquote{Machine learning materials physics:
  Deep neural networks trained on elastic free energy data from martensitic
  microstructures predict homogenized stress fields with high accuracy,}
  \emph{arXiv preprint arXiv:1901.00524}, 2019.

\bibitem[{Fritzen and Kunc(2018)}]{fritzen2018two}
Fritzen, F., and Kunc, O., \enquote{Two-stage data-driven homogenization for
  nonlinear solids using a reduced order model,} \emph{European Journal of
  Mechanics-A/Solids}, Vol.~69, 2018, pp. 201--220.

\bibitem[{Yuan et~al.(2018)Yuan, Paradiso, Meredig, and
  Niezgoda}]{yuan2018machine}
Yuan, M., Paradiso, S., Meredig, B., and Niezgoda, S.~R., \enquote{Machine
  Learning--Based Reduce Order Crystal Plasticity Modeling for ICME
  Applications,} \emph{Integrating Materials and Manufacturing Innovation},
  Vol.~7, No.~4, 2018, pp. 214--230.

\bibitem[{Oishi and Yagawa(2017)}]{oishi2017computational}
Oishi, A., and Yagawa, G., \enquote{Computational mechanics enhanced by deep
  learning,} \emph{Computer Methods in Applied Mechanics and Engineering}, Vol.
  327, 2017, pp. 327--351.

\bibitem[{Capuano and Rimoli(2019)}]{capuano2019smart}
Capuano, G., and Rimoli, J.~J., \enquote{Smart finite elements: A novel machine
  learning application,} \emph{Computer Methods in Applied Mechanics and
  Engineering}, Vol. 345, 2019, pp. 363--381.

\bibitem[{Wei et~al.(2018)Wei, Zhao, Rong, and Bao}]{wei2018predicting}
Wei, H., Zhao, S., Rong, Q., and Bao, H., \enquote{Predicting the effective
  thermal conductivities of composite materials and porous media by machine
  learning methods,} \emph{International Journal of Heat and Mass Transfer},
  Vol. 127, 2018, pp. 908--916.

\bibitem[{Rong et~al.(2019)Rong, Wei, Huang, and Bao}]{rong2019predicting}
Rong, Q., Wei, H., Huang, X., and Bao, H., \enquote{Predicting the effective
  thermal conductivity of composites from cross sections images using deep
  learning methods,} \emph{Composites Science and Technology}, Vol. 184, 2019,
  p. 107861.

\bibitem[{Koeppe et~al.(2020)Koeppe, Bamer, and
  Markert}]{koeppe2020intelligent}
Koeppe, A., Bamer, F., and Markert, B., \enquote{An intelligent nonlinear meta
  element for elastoplastic continua: deep learning using a new
  Time-distributed Residual U-Net architecture,} \emph{Computer Methods in
  Applied Mechanics and Engineering}, Vol. 366, 2020, p. 113088.

\bibitem[{Yang et~al.(2020{\natexlab{a}})Yang, Qiu, Xiang, Tang, and
  Guo}]{yang2020exploring}
Yang, H., Qiu, H., Xiang, Q., Tang, S., and Guo, X., \enquote{Exploring
  Elastoplastic Constitutive Law of Microstructured Materials Through
  Artificial Neural Network—A Mechanistic-Based Data-Driven Approach,}
  \emph{Journal of Applied Mechanics}, Vol.~87, No.~9, 2020{\natexlab{a}}.

\bibitem[{Zhang and Mohr(2020)}]{zhang2020using}
Zhang, A., and Mohr, D., \enquote{Using neural networks to represent von Mises
  plasticity with isotropic hardening,} \emph{International Journal of
  Plasticity}, Vol. 132, 2020, p. 102732.

\bibitem[{Wu et~al.(2020{\natexlab{b}})Wu, Kilingar, Noels
  et~al.}]{wu2020recurrent}
Wu, L., Kilingar, N.~G., Noels, L., et~al., \enquote{A recurrent neural
  network-accelerated multi-scale model for elasto-plastic heterogeneous
  materials subjected to random cyclic and non-proportional loading paths,}
  \emph{Computer Methods in Applied Mechanics and Engineering}, Vol. 369,
  2020{\natexlab{b}}, p. 113234.

\bibitem[{Stoffel et~al.(2018)Stoffel, Bamer, and
  Markert}]{stoffel2018artificial}
Stoffel, M., Bamer, F., and Markert, B., \enquote{Artificial neural networks
  and intelligent finite elements in non-linear structural mechanics,}
  \emph{Thin-Walled Structures}, Vol. 131, 2018, pp. 102--106.

\bibitem[{Stoffel et~al.(2019)Stoffel, Bamer, and Markert}]{stoffel2019neural}
Stoffel, M., Bamer, F., and Markert, B., \enquote{Neural network based
  constitutive modeling of nonlinear viscoplastic structural response,}
  \emph{Mechanics Research Communications}, Vol.~95, 2019, pp. 85--88.

\bibitem[{Ghavamian and Simone(2019)}]{ghavamian2019accelerating}
Ghavamian, F., and Simone, A., \enquote{Accelerating multiscale finite element
  simulations of history-dependent materials using a recurrent neural network,}
  \emph{Computer Methods in Applied Mechanics and Engineering}, Vol. 357, 2019,
  p. 112594.

\bibitem[{Stoffel et~al.(2020)Stoffel, Bamer, and Markert}]{stoffel2020deep}
Stoffel, M., Bamer, F., and Markert, B., \enquote{Deep convolutional neural
  networks in structural dynamics under consideration of viscoplastic material
  behaviour,} \emph{Mechanics Research Communications}, 2020, p. 103565.

\bibitem[{Shen et~al.(2005)Shen, Chandrashekhara, Breig, and
  Oliver}]{shen2005finite}
Shen, Y., Chandrashekhara, K., Breig, W., and Oliver, L., \enquote{Finite
  element analysis of V-ribbed belts using neural network based hyperelastic
  material model,} \emph{International Journal of Non-Linear Mechanics},
  Vol.~40, No.~6, 2005, pp. 875--890.

\bibitem[{Hambli et~al.(2011)Hambli, Katerchi, and
  Benhamou}]{hambli2011multiscale}
Hambli, R., Katerchi, H., and Benhamou, C.-L., \enquote{Multiscale methodology
  for bone remodelling simulation using coupled finite element and neural
  network computation,} \emph{Biomechanics and Modeling in Mechanobiology},
  Vol.~10, No.~1, 2011, pp. 133--145.

\bibitem[{Settgast et~al.(2020)Settgast, H{\"u}tter, Kuna, and
  Abendroth}]{settgast2020hybrid}
Settgast, C., H{\"u}tter, G., Kuna, M., and Abendroth, M., \enquote{A hybrid
  approach to simulate the homogenized irreversible elastic--plastic
  deformations and damage of foams by neural networks,} \emph{International
  Journal of Plasticity}, Vol. 126, 2020, p. 102624.

\bibitem[{Fern{\'a}ndez et~al.(2020)Fern{\'a}ndez, Rezaei, Mianroodi, Fritzen,
  and Reese}]{fernandez2020application}
Fern{\'a}ndez, M., Rezaei, S., Mianroodi, J.~R., Fritzen, F., and Reese, S.,
  \enquote{Application of artificial neural networks for the prediction of
  interface mechanics: a study on grain boundary constitutive behavior,}
  \emph{Advanced Modeling and Simulation in Engineering Sciences}, Vol.~7,
  No.~1, 2020, pp. 1--27.

\bibitem[{Systemes(2014)}]{systemes2014abaqus}
Systemes, D., \enquote{Abaqus User Subroutines Reference Guide, Version 6.14,}
  \emph{Dassault Systemes Simulia Corp., Providence, RI, USA}, 2014.

\bibitem[{Raissi et~al.(2019)Raissi, Perdikaris, and
  Karniadakis}]{raissi2019physics}
Raissi, M., Perdikaris, P., and Karniadakis, G.~E., \enquote{Physics-informed
  neural networks: A deep learning framework for solving forward and inverse
  problems involving nonlinear partial differential equations,} \emph{Journal
  of Computational Physics}, Vol. 378, 2019, pp. 686--707.

\bibitem[{Tao et~al.(2020)Tao, Liu, Du, and Yu}]{tao2020physics}
Tao, F., Liu, X., Du, H., and Yu, W., \enquote{Physics-informed artificial
  neural network approach for axial compression buckling analysis of
  thin-walled cylinder,} \emph{AIAA Journal}, Vol.~58, No.~6, 2020, pp.
  2737--2747.

\bibitem[{Zhang et~al.(2020)Zhang, Liu, and Sun}]{zhang2020PINN}
Zhang, R., Liu, Y., and Sun, H., \enquote{Physics-informed multi-LSTM networks
  for metamodeling of nonlinear structures,} \emph{Computer Methods in Applied
  Mechanics and Engineering}, Vol. 369, 2020, p. 113226.

\bibitem[{Masi et~al.(2020)Masi, Stefanou, Vannucci, and
  Maffi-Berthier}]{masi2020thermodynamics}
Masi, F., Stefanou, I., Vannucci, P., and Maffi-Berthier, V.,
  \enquote{Thermodynamics-based Artificial Neural Networks for constitutive
  modeling,} \emph{arXiv preprint arXiv:2005.12183}, 2020.

\bibitem[{Liu et~al.(2018)Liu, Fleming, and Liu}]{liu2018microstructural}
Liu, Z., Fleming, M., and Liu, W.~K., \enquote{Microstructural material
  database for self-consistent clustering analysis of elastoplastic strain
  softening materials,} \emph{Computer Methods in Applied Mechanics and
  Engineering}, Vol. 330, 2018, pp. 547--577.

\bibitem[{Rocha et~al.(2020{\natexlab{b}})Rocha, van~der Meer, and
  Sluys}]{rocha2020adaptive}
Rocha, I., van~der Meer, F., and Sluys, L., \enquote{An adaptive domain-based
  POD/ECM hyper-reduced modeling framework without offline training,}
  \emph{Computer Methods in Applied Mechanics and Engineering}, Vol. 358,
  2020{\natexlab{b}}, p. 112650.

\bibitem[{Michel and Suquet(2010)}]{michel2010non}
Michel, J.-C., and Suquet, P., \enquote{Non-uniform transformation field
  analysis: a reduced model for multiscale non-linear problems in solid
  mechanics,} \emph{Multiscale Modeling In Solid Mechanics: Computational
  Approaches}, World Scientific, 2010, pp. 159--206.

\bibitem[{Liu and Wang(2019)}]{liu2019multi}
Liu, D., and Wang, Y., \enquote{Multi-fidelity physics-constrained neural
  network and its application in materials modeling,} \emph{Journal of
  Mechanical Design}, Vol. 141, No.~12, 2019.

\bibitem[{Yoo et~al.(2020)Yoo, Bacarreza, and Aliabadi}]{yoo2020novel}
Yoo, K., Bacarreza, O., and Aliabadi, M.~F., \enquote{A novel multi-fidelity
  modelling-based framework for reliability-based design optimisation of
  composite structures,} \emph{Engineering with Computers}, 2020.
\newblock \urlprefix\url{https://doi.org/10.1007/s00366-020-01084-x}.

\bibitem[{Tian et~al.(2020)Tian, Ma, Li, Lin, Wang, and Waas}]{tian2020multi}
Tian, K., Ma, X., Li, Z., Lin, S., Wang, B., and Waas, A.~M., \enquote{A
  multi-fidelity competitive sampling method for surrogate-based stacking
  sequence optimization of composite shells with multiple cutouts,}
  \emph{International Journal of Solids and Structures}, Vol. 193, 2020, pp.
  1--12.

\bibitem[{Meng and Karniadakis(2020)}]{meng2020composite}
Meng, X., and Karniadakis, G.~E., \enquote{A composite neural network that
  learns from multi-fidelity data: Application to function approximation and
  inverse PDE problems,} \emph{Journal of Computational Physics}, Vol. 401,
  2020, p. 109020.

\bibitem[{Wang and Shan(2006)}]{wang2007review}
Wang, G.~G., and Shan, S., \enquote{Review of Metamodeling Techniques in
  Support of Engineering Design Optimization,} \emph{Journal of Mechanical
  Design}, Vol. 129, No.~4, 2006, pp. 370--380.

\bibitem[{Yao et~al.(2020)Yao, Gao, and Liu}]{yao2020fea}
Yao, H., Gao, Y., and Liu, Y., \enquote{FEA-Net: A physics-guided data-driven
  model for efficient mechanical response prediction,} \emph{Computer Methods
  in Applied Mechanics and Engineering}, Vol. 363, 2020, p. 112892.

\bibitem[{Gao et~al.(2020)Gao, Yao, Wei, and Liu}]{gao2020physics}
Gao, Y., Yao, H., Wei, H., and Liu, Y., \enquote{Physics-based Deep Learning
  for Probabilistic Fracture Analysis of Composite Materials,} \emph{AIAA
  Scitech 2020 Forum}, 2020, p. 1860.

\bibitem[{Zobeiry et~al.(2020)Zobeiry, Reiner, and Vaziri}]{zobeiry2020theory}
Zobeiry, N., Reiner, J., and Vaziri, R., \enquote{Theory-guided machine
  learning for damage characterization of composites,} \emph{Composite
  Structures}, 2020, p. 112407.

\bibitem[{Simpson et~al.(2004)Simpson, Booker, Ghosh, Giunta, Koch, and
  Yang}]{simpson2004approximation}
Simpson, T.~W., Booker, A.~J., Ghosh, D., Giunta, A.~A., Koch, P.~N., and Yang,
  R.-J., \enquote{Approximation methods in multidisciplinary analysis and
  optimization: a panel discussion,} \emph{Structural and multidisciplinary
  optimization}, Vol.~27, No.~5, 2004, pp. 302--313.

\bibitem[{Viana et~al.(2014)Viana, Simpson, Balabanov, and
  Toropov}]{viana2014special}
Viana, F.~A., Simpson, T.~W., Balabanov, V., and Toropov, V., \enquote{Special
  section on multidisciplinary design optimization: metamodeling in
  multidisciplinary design optimization: how far have we really come?}
  \emph{AIAA journal}, Vol.~52, No.~4, 2014, pp. 670--690.

\bibitem[{Shu et~al.(2017)Shu, Jiang, Wan, Zhou, Shao, and
  Zhang}]{shu2017metamodel}
Shu, L., Jiang, P., Wan, L., Zhou, Q., Shao, X., and Zhang, Y.,
  \enquote{Metamodel-based design optimization employing a novel sequential
  sampling strategy,} \emph{Engineering Computations}, 2017.

\bibitem[{Jiang et~al.(2020)Jiang, Zhou, and Shao}]{jiang2020surrogate}
Jiang, P., Zhou, Q., and Shao, X., \enquote{Surrogate-Model-Based Design and
  Optimization,} \emph{Surrogate Model-Based Engineering Design and
  Optimization}, Springer, 2020, pp. 135--236.

\bibitem[{Muc and Gurba(2001)}]{muc2001genetic}
Muc, A., and Gurba, W., \enquote{Genetic algorithms and finite element analysis
  in optimization of composite structures,} \emph{Composite Structures},
  Vol.~54, No. 2-3, 2001, pp. 275--281.

\bibitem[{Chen and Cheng(2010)}]{chen2010data}
Chen, T.-Y., and Cheng, Y.-L., \enquote{Data-mining assisted structural
  optimization using the evolutionary algorithm and neural network,}
  \emph{Engineering Optimization}, Vol.~42, No.~3, 2010, pp. 205--222.

\bibitem[{D{\'\i}az et~al.(2016)D{\'\i}az, Montoya, and
  Hern{\'a}ndez}]{diaz2016efficient}
D{\'\i}az, J., Montoya, M.~C., and Hern{\'a}ndez, S., \enquote{Efficient
  methodologies for reliability-based design optimization of composite panels,}
  \emph{Advances in Engineering Software}, Vol.~93, 2016, pp. 9--21.

\bibitem[{Truong et~al.(2020)Truong, Lee, and Lee}]{truong2020artificial}
Truong, T.~T., Lee, S., and Lee, J., \enquote{An artificial neural
  network-differential evolution approach for optimization of bidirectional
  functionally graded beams,} \emph{Composite Structures}, Vol. 233, 2020, p.
  111517.

\bibitem[{Chen and Gu(2020)}]{chen2020generative}
Chen, C.-T., and Gu, G.~X., \enquote{Generative deep neural networks for
  inverse materials design using backpropagation and active learning,}
  \emph{Advanced Science}, Vol.~7, No.~5, 2020, p. 1902607.

\bibitem[{Bisagni and Lanzi(2002)}]{bisagni2002post}
Bisagni, C., and Lanzi, L., \enquote{Post-buckling optimisation of composite
  stiffened panels using neural networks,} \emph{Composite Structures},
  Vol.~58, No.~2, 2002, pp. 237--247.

\bibitem[{Abouhamze and Shakeri(2007)}]{abouhamze2007multi}
Abouhamze, M., and Shakeri, M., \enquote{Multi-objective stacking sequence
  optimization of laminated cylindrical panels using a genetic algorithm and
  neural networks,} \emph{Composite Structures}, Vol.~81, No.~2, 2007, pp.
  253--263.

\bibitem[{Mar{\'\i}n et~al.(2012)Mar{\'\i}n, Trias, Badall{\'o}, Rus, and
  Mayugo}]{marin2012optimization}
Mar{\'\i}n, L., Trias, D., Badall{\'o}, P., Rus, G., and Mayugo, J.,
  \enquote{Optimization of composite stiffened panels under mechanical and
  hygrothermal loads using neural networks and genetic algorithms,}
  \emph{Composite Structures}, Vol.~94, No.~11, 2012, pp. 3321--3326.

\bibitem[{Fu et~al.(2015)Fu, Ricci, and Bisagni}]{fu2015minimum}
Fu, X., Ricci, S., and Bisagni, C., \enquote{Minimum-weight design for three
  dimensional woven composite stiffened panels using neural networks and
  genetic algorithms,} \emph{Composite Structures}, Vol. 134, 2015, pp.
  708--715.

\bibitem[{Yazdi et~al.(2016)Yazdi, Rostami, and
  Kolahdooz}]{yazdi2016optimization}
Yazdi, M.~S., Rostami, S.~L., and Kolahdooz, A., \enquote{Optimization of
  geometrical parameters in a specific composite lattice structure using neural
  networks and ABC algorithm,} \emph{Journal of Mechanical Science and
  Technology}, Vol.~30, No.~4, 2016, pp. 1763--1771.

\bibitem[{Garmsiri and Jalal(2014)}]{garmsiri2014multiobjective}
Garmsiri, K., and Jalal, M., \enquote{Multiobjective optimization of composite
  cylindrical shells for strength and frequency using genetic algorithm and
  neural networks,} \emph{Science and Engineering of Composite Materials},
  Vol.~21, No.~4, 2014, pp. 529--536.

\bibitem[{Gomes et~al.(2017)Gomes, Diniz, da~Cunha, and
  Ancelotti}]{gomes2017design}
Gomes, G.~F., Diniz, C.~A., da~Cunha, S.~S., and Ancelotti, A.~C.,
  \enquote{Design optimization of composite prosthetic tubes using ga-ann
  algorithm considering tsai-wu failure criteria,} \emph{Journal of Failure
  Analysis and Prevention}, Vol.~17, No.~4, 2017, pp. 740--749.

\bibitem[{Abueidda et~al.(2020)Abueidda, Koric, and
  Sobh}]{abueidda2020topology}
Abueidda, D.~W., Koric, S., and Sobh, N.~A., \enquote{Topology optimization of
  2D structures with nonlinearities using deep learning,} \emph{Computers \&
  Structures}, Vol. 237, 2020, p. 106283.

\bibitem[{Wilt et~al.(2020)Wilt, Yang, and Gu}]{wilt2020accelerating}
Wilt, J.~K., Yang, C., and Gu, G.~X., \enquote{Accelerating Auxetic
  Metamaterial Design with Deep Learning,} \emph{Advanced Engineering
  Materials}, Vol.~22, No.~5, 2020, p. 1901266.

\bibitem[{Nie et~al.(2020)Nie, Jiang, and Kara}]{nie2020stress}
Nie, Z., Jiang, H., and Kara, L.~B., \enquote{Stress field prediction in
  cantilevered structures using convolutional neural networks,} \emph{Journal
  of Computing and Information Science in Engineering}, Vol.~20, No.~1, 2020.

\bibitem[{Banga et~al.(2018)Banga, Gehani, Bhilare, Patel, and
  Kara}]{banga20183d}
Banga, S., Gehani, H., Bhilare, S., Patel, S., and Kara, L., \enquote{3d
  topology optimization using convolutional neural networks,} \emph{arXiv
  preprint arXiv:1808.07440}, 2018.

\bibitem[{Gu et~al.(2018)Gu, Chen, and Buehler}]{gu2018novo}
Gu, G.~X., Chen, C.-T., and Buehler, M.~J., \enquote{De novo composite design
  based on machine learning algorithm,} \emph{Extreme Mechanics Letters},
  Vol.~18, 2018, pp. 19--28.

\bibitem[{Yildiz et~al.(2003)Yildiz, {\"O}zt{\"u}rk, Kaya, and
  {\"O}zt{\"u}rk}]{yildiz2003integrated}
Yildiz, A., {\"O}zt{\"u}rk, N., Kaya, N., and {\"O}zt{\"u}rk, F.,
  \enquote{Integrated optimal topology design and shape optimization using
  neural networks,} \emph{Structural and Multidisciplinary Optimization},
  Vol.~25, No.~4, 2003, pp. 251--260.

\bibitem[{Wang et~al.(2020)Wang, Yeo, Su, Wang, and Abdalla}]{wang2020data}
Wang, D., Yeo, S.-Y., Su, Z., Wang, Z.-P., and Abdalla, M.~M.,
  \enquote{Data-driven streamline stiffener path optimization (SSPO) for sparse
  stiffener layout design of non-uniform curved grid-stiffened composite (NCGC)
  structures,} \emph{Computer Methods in Applied Mechanics and Engineering},
  Vol. 365, 2020, p. 113001.

\bibitem[{Xiang et~al.(2014)Xiang, Xiang, and Wu}]{xiang2014prediction}
Xiang, K.-L., Xiang, P.-Y., and Wu, Y.-P., \enquote{Prediction of the fatigue
  life of natural rubber composites by artificial neural network approaches,}
  \emph{Materials \& Design}, Vol.~57, 2014, pp. 180--185.

\bibitem[{Balokas et~al.(2018)Balokas, Czichon, and Rolfes}]{balokas2018neural}
Balokas, G., Czichon, S., and Rolfes, R., \enquote{Neural network assisted
  multiscale analysis for the elastic properties prediction of 3D braided
  composites under uncertainty,} \emph{Composite Structures}, Vol. 183, 2018,
  pp. 550--562.

\bibitem[{Han et~al.(2020)Han, Gao, Fleming, Xu, Xie, Meng, and
  Liu}]{han2020efficient}
Han, X., Gao, J., Fleming, M., Xu, C., Xie, W., Meng, S., and Liu, W.~K.,
  \enquote{Efficient multiscale modeling for woven composites based on
  self-consistent clustering analysis,} \emph{Computer Methods in Applied
  Mechanics and Engineering}, Vol. 364, 2020, p. 112929.

\bibitem[{Shields and Aakash(2019)}]{UQreport}
Shields, N.~D., and Aakash, B., \enquote{Uncertainty quantification in
  computational solid and structural materials modeling,} Tech. rep., Johns
  Hopkins University, Baltimore, MD, 2019.

\bibitem[{Acar(2020)}]{acar2020machine}
Acar, P., \enquote{Machine Learning Reinforced Crystal Plasticity Modeling
  under Experimental Uncertainty,} \emph{AIAA Scitech 2020 Forum}, 2020, p.
  1152.

\bibitem[{Gomes et~al.(2011)Gomes, Awruch, and Lopes}]{gomes2011reliability}
Gomes, H.~M., Awruch, A.~M., and Lopes, P. A.~M., \enquote{Reliability based
  optimization of laminated composite structures using genetic algorithms and
  Artificial Neural Networks,} \emph{Structural Safety}, Vol.~33, No.~3, 2011,
  pp. 186--195.

\bibitem[{Song et~al.(2017)Song, Fei, Wen, and Bai}]{song2017multi}
Song, L.-K., Fei, C.-W., Wen, J., and Bai, G.-C., \enquote{Multi-objective
  reliability-based design optimization approach of complex structure with
  multi-failure modes,} \emph{Aerospace Science and Technology}, Vol.~64, 2017,
  pp. 52--62.

\bibitem[{Artero-Guerrero et~al.(2018)Artero-Guerrero, Pernas-S{\'a}nchez,
  Mart{\'\i}n-Montal, Varas, and L{\'o}pez-Puente}]{artero2018influence}
Artero-Guerrero, J., Pernas-S{\'a}nchez, J., Mart{\'\i}n-Montal, J., Varas, D.,
  and L{\'o}pez-Puente, J., \enquote{The influence of laminate stacking
  sequence on ballistic limit using a combined Experimental/FEM/Artificial
  Neural Networks (ANN) methodology,} \emph{Composite Structures}, Vol. 183,
  2018, pp. 299--308.

\bibitem[{Ang et~al.(2018)Ang, Majid, Nor, Yaacob, and Ridzuan}]{ang2018first}
Ang, J., Majid, M.~A., Nor, A.~M., Yaacob, S., and Ridzuan, M.,
  \enquote{First-ply failure prediction of glass/epoxy composite pipes using an
  artificial neural network model,} \emph{Composite Structures}, Vol. 200,
  2018, pp. 579--588.

\bibitem[{El~Said and Hallett(2018)}]{el2018multiscale}
El~Said, B., and Hallett, S.~R., \enquote{Multiscale surrogate modelling of the
  elastic response of thick composite structures with embedded defects and
  features,} \emph{Composite Structures}, Vol. 200, 2018, pp. 781--798.

\bibitem[{Karpatne et~al.(2017)Karpatne, Watkins, Read, and
  Kumar}]{karpatne2017physics}
Karpatne, A., Watkins, W., Read, J., and Kumar, V., \enquote{Physics-guided
  neural networks (pgnn): An application in lake temperature modeling,}
  \emph{arXiv preprint arXiv:1710.11431}, 2017.

\bibitem[{Yang et~al.(2020{\natexlab{b}})Yang, Kim, Ryu, and
  Gu}]{yang2020prediction}
Yang, C., Kim, Y., Ryu, S., and Gu, G.~X., \enquote{Prediction of composite
  microstructure stress-strain curves using convolutional neural networks,}
  \emph{Materials \& Design}, Vol. 189, 2020{\natexlab{b}}, p. 108509.

\end{thebibliography}

\end{document}